\documentclass[3p,times,sort&compress]{elsarticle}
\usepackage{ecrc_own}
\usepackage{lineno,hyperref}
\modulolinenumbers[5]

\journal{Journal of \LaTeX\ Templates}

\usepackage{graphicx}
\usepackage{dcolumn}
\usepackage{bm}
\usepackage{amsmath}
\usepackage{amssymb}
\usepackage{amsfonts}
\usepackage{latexsym}
\usepackage[T1]{fontenc}

\volume{00}

\firstpage{1}

\runauth{Sebastian Bleser et al.}

\usepackage[figuresright]{rotating}

\usepackage[version=3]{mhchem}

\begin{document}

\begin{frontmatter}

\title{Has the neutral double hypernucleus \ce{_{$\Lambda\Lambda$}^{4}n} been observed?}

\author[myaddress01]{Sebastian~Bleser}
\author[myaddress01]{Michael~B\"olting}
\author[myaddress02]{Theodoros~Gaitanos}
\author[myaddress01,myaddress03]{Josef~Pochodzalla\corref{mycorrespondingauthor}}
\cortext[mycorrespondingauthor]{Corresponding author}
\ead{pochodza@uni-mainz.de}
\ead[url]{https://www.hi-mainz.de}
\author[myaddress01]{Falk~Schupp}
\author[myaddress01]{Marcell~Steinen}

\address[myaddress01]{Helmholtz Institute Mainz, Johannes Gutenberg University, 55099 Mainz, Germany.}
\address[myaddress02]{Aristotle University of Thessaloniki, 54124 Thessalon\'{\i}ki, Greece.}
\address[myaddress03]{Institute for Nuclear Physics, Johannes Gutenberg University, 55099 Mainz, Germany.}

\begin{abstract}
The BNL-AGS E906 experiment was the first fully electronic experiment to produce and study double hypernuclei with large statistics. Two dominant structures were observed in the correlated $\pi^-$--$\pi^-$ momentum matrix at (p$_{\pi-H}$,p$_{\pi-L}$) = (133,114)\,MeV/{\em{c}} and at (114,104)\,MeV/{\em{c}}. In this work we argue that
the interpretation of the structure at (133,114)\,MeV/{\em{c}} in terms of $^3_{\Lambda}$H+$^4_{\Lambda}$H pairs
is questionable. We show, that neither a scenario where these singe-$\Lambda$ hypernuclei are produced after capture of a stopped $\Xi^-$ by a $^9$Be nucleus nor interactions of energetic $\Xi^-$ with $^9$Be nuclei in the target material can produce a sufficient amount of such twin pairs. We have therefore explored the conjecture of Avraham Gal that decays of the \ce{_{$\Lambda\Lambda$}^{4}n} may be responsible for the observed structure. Indeed, the inclusion of \ce{_{$\Lambda\Lambda$}^{4}n} with a two-body $\pi^-$ branching ratio of 50\% in a statistical multifragmentation model allows to describe the E906 data remarkably well. On the other hand, a bound $^{3}_{\Lambda}$n nucleus would cause a striking structure in the momentum correlation matrix which is clearly inconsistent with the observation of E906.
\end{abstract}
\begin{keyword}
\texttt{hypernuclei \sep statistical decay model}
\end{keyword}
\end{frontmatter}


\section{Introduction}
\label{sec:intro}

Recently, a candidate for a weakly unbound resonant tetraneutron was observed in the missing-mass spectrum measured in the double-charge-exchange reaction $^4$He($^8$He,$^8$Be) at the Radioactive Isotope Beam Factory of RIKEN \cite{PhysRevLett.116.052501}. Prior to this experiment, first positive indications of such a nucleus were found in 2002 at GANIL employing the breakup of $^{14}$Be nuclei \cite{PhysRevC.65.044006}. In the hypernuclear sector, the HypHI Collaboration investigated peripheral $^6$Li+$^{12}$C interactions at GSI and found intriguing enhancements in the d+${\pi}^{-}$ and t+${\pi}^{-}$ invariant mass. These enhancements were interpreted as weak decays of neutral $_{\Lambda}^{3}$n hypernuclei \cite{PhysRevC.88.041001}.

\begin{figure*}[t]
\centering
\includegraphics[width=0.48\textwidth]{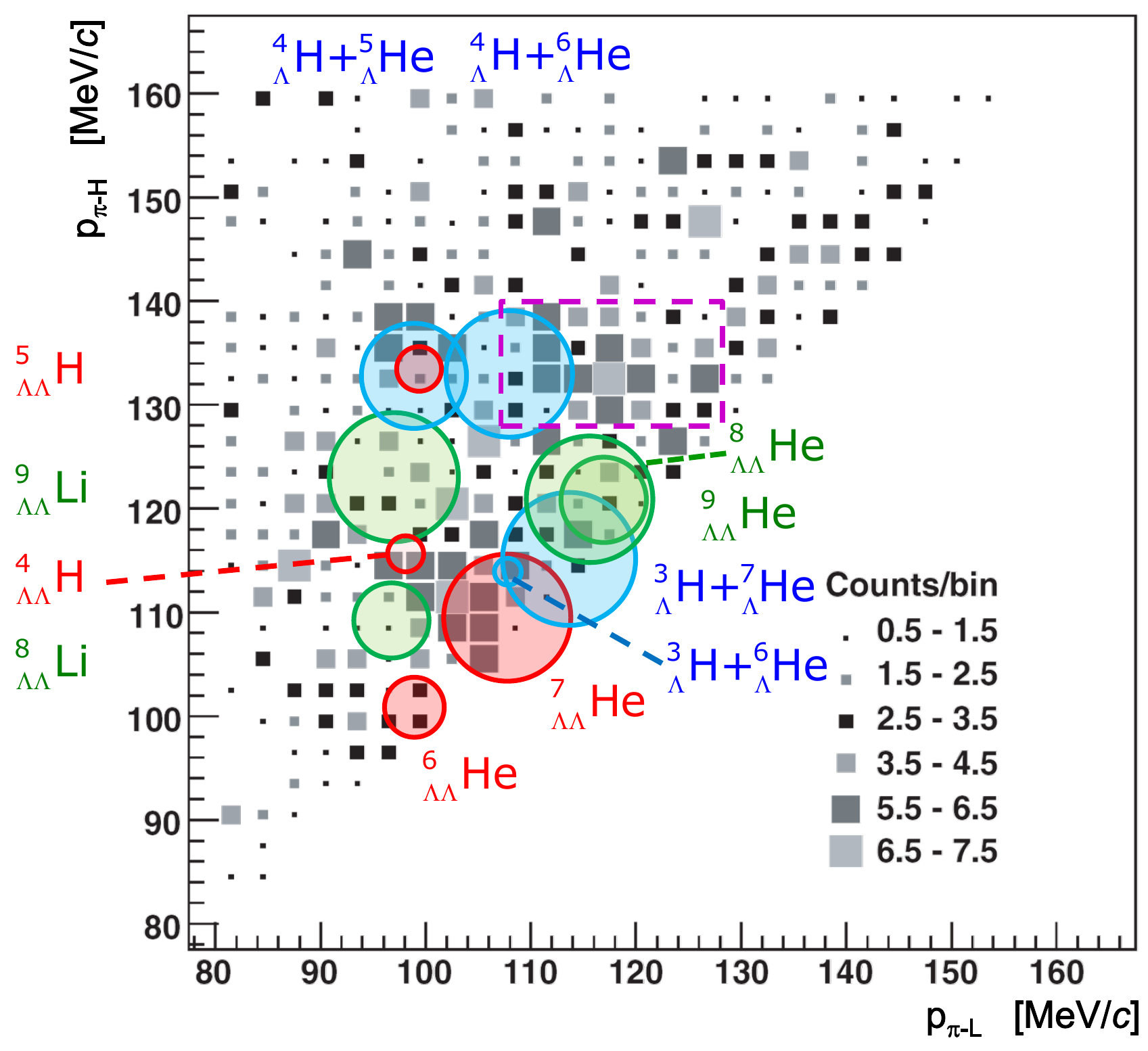}
\includegraphics[width=0.51\textwidth]{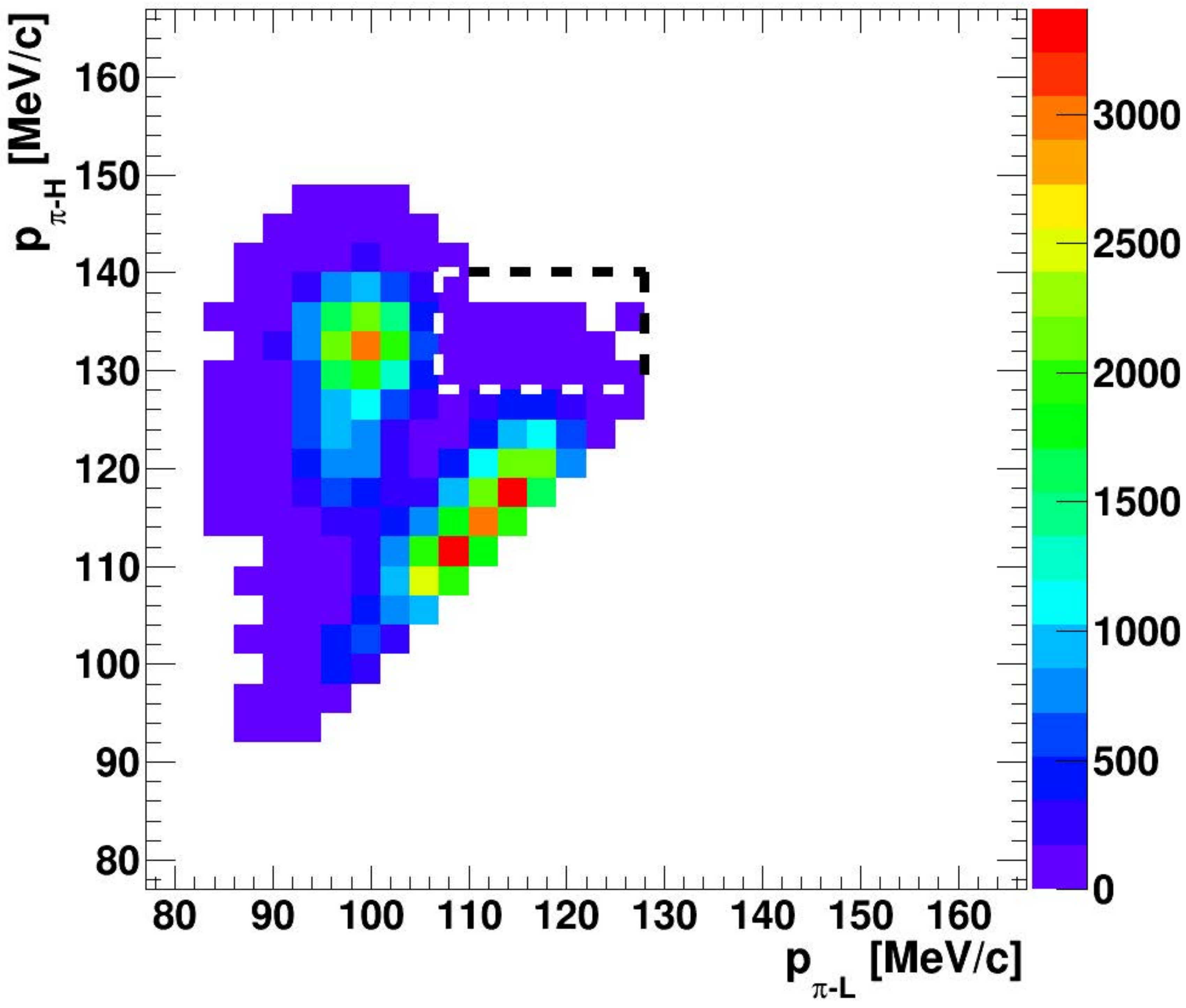}
\caption{Left: Momenta of two correlated pions measured by the E906 collaboration (grey squares).
Overlaid are {\em production} yields of various twin pairs (blue), double hypernuclei with a stable ground state only (red) and double hypernuclei with one or more bound excited states (green). The areas of the circles are proportional to the production yields predicted by the statistical multifragmentation model (SMM), the center of the circles are given by the $\pi^-$ momenta for two-body decays. Pionic decay probabilities are not included here. The neutral hypernuclei $^{3}_{\Lambda}$n and \ce{_{$\Lambda\Lambda$}^{4}n} were not considered in these simulations.
Right: SMM production probabilities are folded with the pionic two-body weak decay branching ratios (Tab.~\ref{tab:double:pipi}) and the momentum resolution of the E906 experiment. The box marks the region of the enhancement seen by the E906 experiment.}
\label{fig:double:E906_corr}
\end{figure*}

From the theoretical point of view, the tetraneutron is rather controversial.
Several theoretical models exclude the existence of a bound or narrow resonant tetraneutron
\cite{PhysRevLett.90.252501,0954-3899-29-2-102,0954-3899-29-10-309,PhysRevC.93.044004,PhysRevLett.119.032501}.
On the other side, calculations of Lashko and Filippov \cite{Lashko2008}, Shirokov {\em et al.} \cite{PhysRevLett.117.182502,PhysRevLett.121.099901} as well as
Gandolfi {\em et al.} \cite{PhysRevLett.118.232501} support a slightly unbound tetraneutron with a resonance energy in agreement with the RIKEN measurement. In the case of the neutral $_{\Lambda}^{3}$n hypernucleus, the theoretical position is clear: based on our present understanding of strange nuclei, this nucleus should not be bound \cite{0305-4616-13-5-002,GAL201493,PhysRevC.89.061302,PhysRevC.89.057001,1674-1137-41-7-074102}.

In view of this challenging situation -- experimentally as well as theoretically -- it is surely justified to explore the possibility of other bound or resonant neutral baryonic systems as well. Of particular interest are nuclear systems with two $\Lambda$ hyperons. The onset of binding in light double $\Lambda$-hypernuclei has a considerable impact on our understanding of three-body forces in dense baryonic matter. For charged double hypernuclei, the theoretical situation is unclear for the A = 4 \ce{_{$\Lambda\Lambda$}^{4}H} nucleus
\cite{PhysRevLett.89.142504,PhysRevLett.89.172502,PhysRevC.67.051001,PhysRevLett.94.202502}, but binding is expected to definitely set in for A = 5 double hypernuclei \cite{FILIKHIN2002491,SweMyint2003,GAL200591}.
Unfortunately, there is no consensus about the stability with respect to either the $\Lambda\Lambda $n
\cite{PhysRevLett.110.012503,PhysRevLett.110.179201,PhysRevLett.110.179202}
nucleus or the \ce{_{$\Lambda\Lambda$}^{4}n} nucleus \cite{1742-6596-569-1-012079,PhysRevC.91.014003,1674-1137-41-7-074102}.

In this work we reconsider data taken by the AGS-E906 experiment which searched for double hypernuclei via their sequential pionic decay \cite{PhysRevLett.87.132504}. Studying different production scenarios, we find hints for a bound \ce{_{$\Lambda\Lambda$}^{4}n} system.

\begin{table*}[tb]
\caption{Accessible decay channels with twin-hypernuclei (upper rows) or $\Lambda\Lambda$-hypernuclei (lower rows) of an excited \ce{_{$\Lambda\Lambda$}^{10}Li}$^*$ hyperfragment which was formed after capture of a stopped $\Xi^-$ by a $^9$Be nucleus. The pion momenta and the branching ratios are given for two-body $\pi^-$ decays.
The neutral nucleus \ce{_{$\Lambda\Lambda$}^{4}n} has not been observed yet and its stability is controversial \cite{PhysRevC.91.014003,1674-1137-41-7-074102}. Also the $^{3}_{\Lambda}$n \cite{PhysRevC.88.041001} needs further confirmation. Both hypernuclei are not included in the {\bf  production probability} calculations shown in Fig.~\ref{fig:double:E906_corr} and column 7.
}
\centering
\begin{tabular}{llcccc|cccc}
\hline
\multicolumn{2}{c}{\bf decay channels}&\multicolumn{2}{c}{\bf $\pi^-$ decay}& \multicolumn{2}{c|}{\bf two-body $\pi^-$ }&   \multicolumn{4}{c}{\bf production probability}    \\
~          &                          &\multicolumn{2}{c}{\bf momenta}     & \multicolumn{2}{c|}{\bf branching}   & ~ & {\bf only} &  {\bf only} &   {\bf  $^3_{\Lambda}$n  +}         \\
~          &                          &\multicolumn{2}{c}{\bf (MeV/{\em{c}})}      & \multicolumn{2}{c|}{\bf ratios}           &  {\bf no neutral} & {\bf $^3_{\Lambda}$n}  & {\bf \ce{_{$\Lambda\Lambda$}^{4}n}}  & {\bf \ce{_{$\Lambda\Lambda$}^{4}n}}\\ \hline

$^{3}_{\Lambda}$n $\rightarrow$ $^{3}$H+$\pi^-$ & $^3_{\Lambda}$H $\rightarrow$ $^{3}$He+$\pi^-$    & 119& 114  & 0.25 &0.249   &   -   &0.0002&  -     & 0.0002    \\

$^{3}_{\Lambda}$n $\rightarrow$ $^{3}$H+$\pi^-$ & $^5_{\Lambda}$He $\rightarrow$ $^{4}$He+p+$\pi^-$  & 119& 99 & 0.25  &0.17          &   -   & 0.0121&   -    &   0.0120         \\

$^{3}_{\Lambda}$n $\rightarrow$ $^{3}$H+$\pi^-$ & $^6_{\Lambda}$He $\rightarrow$ $^{6}$Li+$\pi^-$    & 119 & 108 & 0.25 & 0            &   -   & 0.0012&  -     &   0.0011       \\

$^{3}_{\Lambda}$n $\rightarrow$ $^{3}$H+$\pi^-$& $^7_{\Lambda}$Li $\rightarrow$ $^{7}$Be+$\pi^-$     & 119  &108& 0.25&0.24            &   -   & 0.2421&   -    &   0.2380       \\

$^{3}_{\Lambda}$H $\rightarrow$ $^{3}$He+$\pi^-$& $^3_{\Lambda}$H $\rightarrow$ $^{3}$He+$\pi^-$    & 114 &114 & 0.249  &0.249         &0.0001 & 0.0001& 0.0001 &  0.0001     \\

$^{3}_{\Lambda}$H $\rightarrow$ $^{3}$He+$\pi^-$& $^5_{\Lambda}$He $\rightarrow$ $^{4}$He+p+$\pi^-$    & 114 &99 & 0.249 &0.17          &0.0001 &0.0001&0.0001 &  0.0001     \\

$^{3}_{\Lambda}$H $\rightarrow$ $^{3}$He+$\pi^-$& $^6_{\Lambda}$He $\rightarrow$ $^{6}$Li+$\pi^-$   & 114 &108 & 0.249 &0           &0.0053 & 0.0039& 0.0052&     0.0039       \\

$^{3}_{\Lambda}$H $\rightarrow$ $^{3}$He+$\pi^-$& $^7_{\Lambda}$He $\rightarrow$ $^{7}$Li+$\pi^-$    & 114 & 115 & 0.249  &0.10         &0.1112 & 0.0824&0.1086 &    0.0806       \\

$^{4}_{\Lambda}$H $\rightarrow$ $^{4}$He+$\pi^-$  &$^5_{\Lambda}$He $\rightarrow$ $^{4}$He+p+$\pi^-$   & 133 & 99  & 0.51  &0.17          &0.0716 &0.0533 &0.0701 &     0.0524      \\

$^{4}_{\Lambda}$H $\rightarrow$ $^{4}$He+$\pi^-$& $^6_{\Lambda}$He $\rightarrow$ $^{6}$Li+$\pi^-$    & 133 &108 & 0.51 &0           &0.0981 &0.0727 &0.0960 &     0.0716      \\

\hline
 \ce{_{$\Lambda\Lambda$}^{4}n} $\rightarrow$ $^4_{\Lambda}$H + $\pi^-$&$^4_{\Lambda}$H $\rightarrow$ $^{4}$He+$\pi^-$  &118& 133&0.25&0.51    &    -  &   -    & 0.0227&  0.0170 \\

 \ce{_{$\Lambda\Lambda$}^{4}H} $\rightarrow$ $^4_{\Lambda}$He + $\pi^-$&$^4_{\Lambda}$He $\rightarrow$ $^{4}$Li+$\pi^-$  &117& 98&0.25 &0.19   &0.0087 &0.0066& 0.0085&   0.0063\\

 \ce{_{$\Lambda\Lambda$}^{5}H} $\rightarrow$ $^5_{\Lambda}$He + $\pi^-$& $^5_{\Lambda}$He $\rightarrow$ $^{4}$He+p+$\pi^-$  &134& 99&0.20 &0.17   &0.0209 & 0.0159 & 0.0205& 0.0153  \\

\ce{_{$\Lambda\Lambda$}^{6}He} $\rightarrow$ $^5_{\Lambda}$He + p + $\pi^-$& $^5_{\Lambda}$He $\rightarrow$ $^{4}$He+p+$\pi^-$ &101& 99&0.17&0.17 & 0.0234& 0.0174 & 0.0229& 0.0171 \\

\ce{_{$\Lambda\Lambda$}^{7}He} $\rightarrow$ $^7_{\Lambda}$Li + $\pi^-$& $^7_{\Lambda}$Li $\rightarrow$ $^{7}$Be+$\pi^-$ & 109&108 &0.14 &0.24    &0.1022 &0.0759& 0.1002&   0.0747 \\

\ce{_{$\Lambda\Lambda$}^{8}He} $\rightarrow$ $^8_{\Lambda}$Li + $\pi^-$& $^8_{\Lambda}$Li $\rightarrow$ $^{8}$Be+$\pi^-$ & 116   & 124  &0.13 &0.027   &0.1005 &0.0745& 0.0983&   0.0733  \\
 & &  & 120  &     &0.148 &      &      &       & \\
\ce{_{$\Lambda\Lambda$}^{9}He} $\rightarrow$ $^9_{\Lambda}$Li + $\pi^-$& $^9_{\Lambda}$Li $\rightarrow$ $^{9}$Be+$\pi^-$ & 117&121&0.11&0.1  &0.0229 & 0.0130& 0.0224&  0.0167  \\

 \ce{_{$\Lambda\Lambda$}^{7}Li} $\rightarrow$ $^7_{\Lambda}$Be + $\pi^-$&$^7_{\Lambda}$Be $\rightarrow$ $^{7}$B+$\pi^-$   & 101&96&0.14&0  &0.0002 & 0.0001 &0.0002 &  0.0001  \\

 \ce{_{$\Lambda\Lambda$}^{8}Li} $\rightarrow$ $^8_{\Lambda}$Be + $\pi^-$& $^8_{\Lambda}$Be $\rightarrow$ $^{8}$B+$\pi^-$  &109&97&0.13&0  &0.0369 & 0.0273&0.0359 &  0.0269 \\

 \ce{_{$\Lambda\Lambda$}^{9}Li}  $\rightarrow$ $^9_{\Lambda}$Be + $\pi^-$&$^9_{\Lambda}$Be $\rightarrow$ $^{9}$B+$\pi^-$   &123&97&0.11&0.14  &0.1012 & 0.0753&0.0990 &  0.0748\\
\hline
\end{tabular}
\label{tab:double:pipi}
\end{table*}

\section{The E906 puzzle}
\label{sec:doublehyp:E906puzzle}

In 2001, the AGS-E885/E906 collaboration announced the first 'mass production' of about 30 \ce{_{$\Lambda\Lambda$}^{4}H} events based on
$\sim$10$^4$ stopped $\Xi^-$ \cite{PhysRevLett.87.132504}. The $\Lambda\Lambda$-hypernuclei are identified through the sequential weak decay via $\pi^-$ emission after a (K$^-$,K$^+$) reaction deposited two units of
strangeness in a $^9$Be target. Since the pion momenta are unique fingerprints of weak hypernuclear decays, coincidences between two negative pions helped to trace the sequential decay of a
$\Lambda\Lambda$-nucleus or pairs of single hypernuclei, so called twin hypernuclei.
Two dominant structures were observed in the correlated $\pi^-$--$\pi^-$ momenta at (p$_{\pi-H}$,p$_{\pi-L}$) = (133,114)\,MeV/{\em{c}} and at (114,104)\,MeV/{\em{c}} (see gray histogram in the left part of Fig.~\ref{fig:double:E906_corr}). Another less significant enhancement is also visible at (136,99)\,MeV/{\em{c}}. Here, p$_{\pi-H}$ (p$_{\pi-L}$) denotes the higher (lower) momentum of the two pions.
It is the signal at (133,114)\,MeV/{\em{c}} whose explanation is still obscure and which represents the E906 puzzle discussed in this paper.

To model the E906 experiment, we assume that a $\Xi^-$ is captured by a $^9$Be target nucleus.
We employ a statistical multifragmentation model (SMM) \cite{PhysRevC.76.024909,Lorente2011222} to predict the production of various hyperfragments.
The excitation energy of the initial \ce{_{$\Lambda\Lambda$}^{10}Li^*} system is determined
by the binding energy of the captured $\Xi^-$ at the time of the p$\Xi^- \rightarrow \Lambda\Lambda$ conversion.

In principle we can distinguish between the conversion from an atomic bound state and from a nuclear bound state.
Unfortunately, until now very little experimental data on the nuclear interaction of $\Xi^-$ hyperons is available. Calculations of light $\Xi$ atoms \cite{PhysRevC.59.295}
predict that the conversion of the captured $\Xi^-$ from states
with relatively small binding energies of only $\simeq$ 100\,keV dominates.
The available data suggest a nuclear potential depth around 20\,MeV (see e.g. Refs.~\cite{DOVER1983309,Friedman200789,0954-3899-15-3-008,PhysRevC.61.054603}) which corresponds to typical nuclear binding energies of B$_{\Xi^-} \approx$ 4.5\,MeV.
For $^9$Be nuclei, the capture of a $\Xi^-$ at rest with B$_{\Xi^-}$ = 0\,MeV corresponds to about 29.5\,MeV excitation energy.
A larger initial $\Xi^-$ binding energy results in lower excitation energies.
Indeed, the excitation functions of all important decay channels are rather flat at excitation energies between about 25 and 30\,MeV \cite{Pochodzalla:2010}. As a consequence, the results of the simulations will only depend weakly on whether the $\Xi^-$ conversion happens from an atomic bound state with B$_{\Xi^-} \approx$ 0\,MeV or from a nuclear bound state at B$_{\Xi^-}\approx$ 4\,MeV. If not mentioned otherwise, we assume in the following B$_{\Xi^-}$ = 0\,MeV and thus fix the initial excitation energy of the \ce{_{$\Lambda\Lambda$}^{10}Li^*} system at 29.5\,MeV.

The circles in the left part of Fig.~\ref{fig:double:E906_corr} show the relative probabilities for the {\em production} of bound twin (blue circles) and double hypernuclei (green or red) in the E906 experiment after the capture and conversion of a stopped $\Xi^-$, i.e. $\Xi^-+^9$Be $\rightarrow$ \ce{_{$\Lambda\Lambda$}^{10}Li}$^*$ with B$_{\Xi^-}$ = 0\,MeV. The centers of the circles are fixed by the pion momenta for mesonic two-body decays (cf. Tab.~\ref{tab:double:pipi}). The area of the circles in Fig.~\ref{fig:double:E906_corr} is proportional to the respective production probability.
The ground state and corresponding bound excited states - if they exist - have been added. Weak decay probabilities are not yet taken into account in this graph.
The numbers are also listed in the 7$^{th}$ column in Tab.~\ref{tab:double:pipi}.

In the original publication of E906, the bump at (114,104)\,MeV/{\em{c}} was attributed to pionic decays of the double hypernucleus \ce{_{$\Lambda\Lambda$}^{4}H} \cite{PhysRevLett.87.132504}. Later analyses showed, however, that this interpretation may not be unique \cite{PhysRevC.66.014003,PhysRevC.76.064308,Lorente2011222} and suggested an interpretation of this structure in terms of \ce{_{$\Lambda\Lambda$}^{7}He} instead \cite{PhysRevC.76.064308,Lorente2011222}.

Considering the two pion momenta in the third and fourth column of Tab.~\ref{tab:double:pipi},
the weak cluster of events at (136,99)\,MeV/{\em{c}} may be related to decays of \ce{_{$\Lambda\Lambda$}^{5}H} or \ce{_{$\Lambda$}^{4}H} + \ce{_{$\Lambda$}^{5}He} twins. On first sight,
a larger momentum of p$_{\pi-H} \approx$136\,MeV/{\em{c}} seems high compared to the expected momenta of 133.6 or 132.9\,MeV/{\em{c}}. Important aspects, however, are the limited statistics of the data, the uncertainty of the \ce{_{$\Lambda\Lambda$}^{5}H} binding energy and the accuracy of the absolute momentum calibration of about 1\,MeV/{\em{c}}. Indeed, a fit to the projection of the (133,114)\,MeV/{\em{c}} region onto the p$_{\pi-H}$-axis gave a peak position at 134.5$\pm$1.8\,MeV/{\em{c}} \cite{phdthesisNakano2000}, compared to the expected 132.9\,MeV/{\em{c}}. Thus, the interpretation of the (136,99)\,MeV/{\em{c}} structure in terms of \ce{_{$\Lambda\Lambda$}^{5}H} or of \ce{_{$\Lambda$}^{4}H} + \ce{_{$\Lambda$}^{5}He} decays cannot be ruled out.
Furthermore, \ce{_{$\Lambda$}^{4}H} + \ce{_{$\Lambda$}^{6}He} twins may contribute:
two-body decays of  \ce{_{$\Lambda$}^{6}He} would contribute at p$_{\pi-L} \approx$ 108\,MeV/{\em{c}}.
In early emulsion studies, however, no two-body decays of $^6_{\Lambda}$He were found, although the decays of the neighboring isotopes $^5_{\Lambda}$He and $^7_{\Lambda}$He have been seen \cite{Mayeur2016}, indicating the potential sensitivity of these experiments to $^6_{\Lambda}$He $\rightarrow$ $^6$Li+$\pi^-$ decays. On the other hand, 31 three-body modes $\pi^-$+$^2$H+$^4$He have been observed in total \cite{Gajewski1967105,JURIC19731}. Three-body decays, for which experimental information on the pion range or its kinetic energy is available \cite{Breivik1959,Sacton1960}, indicate pion momenta around 95\,MeV/c. Although we neglect two-body decays of $^6_{\Lambda}$He in the following, the three-body mode may contribute, therefore, to the enhancement at (136,99)\,MeV/{\em{c}}.


The real E906 puzzle is related to the second dominant structure with (p$_{\pi-H}$,p$_{\pi-L}$) = (133,114)\,MeV/{\em{c}}. Obviously, there is no strong decay channel appearing which could produce a bump at (133,114){\,MeV/{\em{c}}}. Because of the two central momenta, the interpretation in terms of $^3_{\Lambda}$H+$^4_{\Lambda}$H twins seemed inevitable (cf. Tab.~\ref{tab:double:pipi}). However, the $^3_{\Lambda}$H+$^4_{\Lambda}$H+t mass lies above the initial mass $m_0= m(\Xi^-)+m(^9Be)$ and can therefore not be produced in the $\Xi^{-}_{stopped}+^9$Be compound nucleus production scheme. Similar intriguing is the fact that statistical decay models usually predict a production of $^4_{\Lambda}$H+$^4_{\Lambda}$H twins which exceeds the $^3_{\Lambda}$H+$^4_{\Lambda}$H production \cite{Lorente2011222}. Including additionally the different branching ratios for two-body $\pi^-$ decays of $\Gamma_{\pi^-+^3\textup{He}}/\Gamma_{total} \approx 0.26$ \cite{PhysRevC.57.1595} and $\Gamma_{\pi^-+^4\textup{He}}/\Gamma_{total} \approx 0.5$ \cite{OUTA1998251c}, the absence of a bump which could be attributed to $^4_{\Lambda}$H+$^4_{\Lambda}$H is even more puzzling.


The right part of Fig.~\ref{fig:double:E906_corr} shows the result of the SMM calculations now taking into account the pionic two-body $\pi^-$ branching ratios and the momentum resolution of 3.5\,{\,MeV/{\em{c}}} of the E906 experiment. The binning of the momentum scale is the same as in the experimental matrix of E906. The applied branching ratios are listed in Tab.~\ref{tab:double:pipi}. A detailed justification of these numbers can be found in Ref.~\cite{phdthesisBleser2018}. The dashed rectangle highlights the region of the enhancement seen by the E906 experiment. Even after weak decays, no accumulation of events in the region of interest is discernible.

\begin{table*}[tb]
\centering
\caption{Probability per produced $\Xi^-$ for different $S$=-2 channels by nuclear $\Xi^-$+$^9$Be reactions and by
stopped $\Xi^-$ hyperons. Here, SH (DH) stands for single (double) hypernucleus, respectively.
In case of the stopped $\Xi^-$ process a capture $\times$ conversion probability for producing excited $\Lambda\Lambda$ nuclear systems of 5\% was taken into account. The probabilities in the last 5 columns are multiplied by 10$^4$.}
\begin{tabular}{lclccccc}
\hline
{\bf process}    & {\bf probability [\%]}     & {\bf model} & {\bf $\Lambda$+$\Lambda$} & {\bf $\Lambda$+SH} & {\bf SH+SH}   &{\bf DH} & {\bf $^3_{\Lambda}$H+$^4_{\Lambda}$H}\\ \hline
$\Xi^-$+$^9$Be  & 2.83& GiBUU                               & 37.2& 20.9& 0.070 &3.23 & 0.016\\ \hline
stopped $\Xi^-$ & 4.65 $\times$ 0.05& SMM  $^{10}_{\Lambda\Lambda}$Li$^*$ & 0.525 & 6.365&6.659 & 9.699 &0 \\ 
\end{tabular}
\label{tab:double:smmbuu}
\end{table*}


As already stressed before, the $^3_{\Lambda}$H+$^4_{\Lambda}$H+t mass lies well above the initial mass $m_0= m(\Xi^-)+m(^9Be)$ and such twins can therefore not be produced in the $\Xi^{-}_{stopped}+^9$Be compound production scheme. Of course, the capture from a nuclear bound state can not result in the production of $^3_{\Lambda}$H+$^4_{\Lambda}$H twins.
In principle, $^3_{\Lambda}$n-$^4_{\Lambda}$H pairs could also produce an enhancement around (133,114){\,MeV/{\em{c}}}.
However, the $^3_{\Lambda}$n+$^4_{\Lambda}$H+$^3$He mass exceeds the available energy as well.
With an energy of 12.6\,MeV below $m_0$, the channel $^4_{\Lambda}$H+$^6_{\Lambda}$He is the most likely twin in the present scenario, followed by the $^4_{\Lambda}$H+$^5_{\Lambda}$He+n decay. Taking into account that no two-body decay of $^6_{\Lambda}$He are observed (see previous discussion) and
given the experimental precision of $\approx$1\,{\,MeV/{\em{c}}} for the momentum calibration in E906 \cite{PhysRevLett.87.132504}, neither the decay of $^{4}_{\Lambda}$H+$^6_{\Lambda}$He with (133,108){\,MeV/{\em{c}}} nor decays of $^{4}_{\Lambda}$H+$^5_{\Lambda}$He pairs with (133,99){\,MeV/{\em{c}}} seem to explain the structure around (133,114){\,MeV/{\em{c}}}.

The experimental plot of E906 contains a total of N$_{total}^{E906}$ = 820 counts in 378 bins. Out of these, N$_{\Box}^{E906}$~=~111 events  are found in the region of interest marked by the rectangle in left panel of Fig.~\ref{fig:double:E906_corr}, giving a raw ratio of N$_{\Box}^{E906}$/N$_{total}^{E906}$ = 0.13.
Note, that the simulations do not include background from free hyperon decays or from multi-body decays, which usually will produce relatively low momentum pion pairs. In order to correct the data for this background, we take the three upper horizontal rows with p$_{\pi-H}\,>$\,152\,MeV/{\em{c}} and
the first three vertical columns with  p$_{\pi-L}\,<\,$~89\,MeV/{\em{c}}
in Fig.~\ref{fig:double:E906_corr} as being representative for an equally distributed background. This resulted in 1.24 background events per bin and a corrected relative signal strength of \~{N}$_{\Box}^{E906}$/\~{N}$_{total}^{E906}$=(111-35)/(820-469) = 0.22$\pm$0.04$_{stat}$. In the SMM results shown in the right panel of Fig.~\ref{fig:double:E906_corr} this strength is a factor of 25 smaller, namely 8.7$\cdot$10$^{-3}$. Obviously, these simulations are far from being able to explain the structure observed by the E906 experiment.

In view of this puzzling situation, a production process different from the ones discussed so far seems to be necessary in order to explain the singular structure at (133,114){\,MeV/{\em{c}}} in terms of $^3_{\Lambda}$H+$^4_{\Lambda}$H twins. In Ref. \cite{PhysRevLett.87.132504} it has already been suggested that nuclear reactions of energetic $\Xi^{-}$ hyperons within the beryllium target may be a possible source of $^3_{\Lambda}$H+$^4_{\Lambda}$H twins. As an alternative, Avraham Gal conjectured in a talk at the HYP-2003 conference that the sequential weak decay
\ce{_{$\Lambda\Lambda$}^{4}n}$\rightarrow \pi^-+^4_{\Lambda}$H (see also Tab.~\ref{tab:double:pipi}) might also cause an enhancement around (133,114)\,MeV/{\em{c}} \cite{Gal2003-private,HYP2003-Pile}. In the following we will explore both suggestions.

\section{Can $\Xi^-$-$^9$Be reactions resolve the puzzle?}

The (K$^-$,K$^+$) reaction employed by E906 produces $\Xi^-$ hyperons with a broad momentum spectrum. Only $\Xi^-$ with momenta below approximately 500{\,MeV/{\em{c}}} can be stopped with high efficiency in a target material. More energetic $\Xi^-$ hyperons will decay in flight or will perform hadronic interactions which eventually may produce hypernuclei as well.

The $\Xi^-$  momentum distribution of E906 was obtained from a fit to the free $\Xi^-$ energy spectrum of E906 \cite{phdthesisNakano2000} and a transformation to a momentum scale. These $\Xi^-$ hyperons were propagated in the beryllium target by means of a GEANT simulation. The in-flight $\Xi^-$ decays represent with 92.5\% the dominating process. In total, 4.65\% of the free $\Xi^-$ are brought to rest in the beryllium target and may be captured to form a double hypernucleus. With 2.83\%, the probability for a $\Xi^-$-$^9$Be reaction is similar to the probability of stopped $\Xi^-$.

In order to estimate the contribution by secondary $\Xi^-$+$^9$Be nuclear interactions to the production of $^3_{\Lambda}$H+$^4_{\Lambda}$H twins, the momentum distribution of interacting $\Xi^-$s was convolved with predictions of the Giessen Boltzmann-Uehling-Uhlenbeck (GiBUU) transport model \cite{Buss20121,GAITANOS2014256}.
Two different YN-interaction models were considered, which are based on the microscopic calculations of the Nijmegen group by Rijken and Yamamoto \cite{Rijken:2006kg,doi:10.1143/PTPS.185.14} and by Fujiwara {\em et al.} \cite{PhysRevC.64.054001} (see Ref.~\cite{GAITANOS2014256} for more details).
While for the Fujiwara model the production of two free $\Lambda$ particles continuously increases, the Rijken model predicts a maximum around an incident energy of 100\,MeV followed by a continuous decline for rising energy. This can be traced back to the different behaviour of the $\Xi$N cross sections: in the Rijken model the elastic channel basically vanishes and the inelastic channel drastically drops down at higher energies. On the other hand, both channels stabilize at a finite value in the Fujiwara description \cite{GAITANOS2014256}. As a consequence, the production of twin hypernuclei is about a factor of 5 larger when using the Fujiwara description. Since we are mainly interested in an upper limit of the $^3_{\Lambda}$H+$^4_{\Lambda}$H twin production, we focus on the results with the Fujiwara model.

Over the whole relevant momentum range, two free $\Lambda$ particles or a free $\Lambda$ plus a single $\Lambda$-hypernucleus ($\Lambda$+SH) are the dominating channels. Only at low momenta, where the stopping probability of $\Xi^-$ hyperons is also large, double hypernuclei (DH) have a similar probability to be produced. The most remarkable fact is that even with the Fujiwara model the probability for twin hypernuclei is well below the per mille level. The production of $^3_{\Lambda}$H+$^4_{\Lambda}$H twins in particular reaches only a probability per interaction of the order of 10$^{-4}$. The first row in Tab.~\ref{tab:double:smmbuu} summarizes the production probabilities for various channels
by $\Xi^-$+$^9$Be interactions.  These probabilities are normalized to the total number of produced $\Xi^-$ in the (K$^-$,K$^+$) reaction.

Concerning the production via stopped $\Xi^-$ hyperons, a comparable rate estimate has to take into account not only the $\Xi^-$ production and stopping probability but also the capture probability, the p$\Xi^-\rightarrow\Lambda\Lambda$ conversion and finally the fragmentation processes. The latter is treated by the statistical multifragmentation model. However, the other two missing pieces -- capture and conversion -- require additional considerations.

Unfortunately, a complete description of the $\Xi^-$ atomic cascade and the p$\Xi^-\rightarrow\Lambda\Lambda$ conversion is not feasible at present and would require further theoretical inputs. However, several theoretical estimates for the capture and sticking probability exist which can be used as a guidance
\cite{YAMADA199579,PhysRevC.56.3216,YAMADA1998385c,HIRATAA1998389c,doi:10.1143/PTP.102.89,doi:10.1143/ptp/91.4.747,doi:10.1143/ptp.117.281}. Existing experimental data from earlier emulsion experiments also provide an estimate for these two factors \cite{AOKI2009191}. Consistent with these estimates, we assume a $\Xi^-$ capture $\times$ p$\Xi^- \rightarrow \Lambda\Lambda$ conversion probability of 5\% and describe the subsequent decay of the excited $\Lambda\Lambda$ pre-fragment by the statistical multifragmentation model \cite{Lorente2011222}.

Comparing both production schemes in Tab.~\ref{tab:double:smmbuu}, it is apparent that
$\Xi^-$+$^9$Be reactions strongly contribute to the production of free $\Lambda$ particles.
The production of pairs of one free $\Lambda$ hyperon with one single $\Lambda$-hypernucleus and of $\Lambda\Lambda$-nuclei is comparable to the yield from stopped $\Xi^-$ reactions as well. However, the production of twin hypernuclei (SH+SH) is about two orders of magnitude less likely compared to the yield of twins or double hypernuclei from the stopped $\Xi^-$ process.
Of course, these estimates might have significant uncertainties caused e.g. by unknown branching ratios or uncertainties in the p$\Xi^- \rightarrow \Lambda\Lambda$-conversion process. But it is also clear from  the numbers in Tab.~\ref{tab:double:smmbuu} that neither stopped $\Xi^-$ hyperons nor energetic $\Xi^-$+$^9$Be interactions are able to produce $^3_{\Lambda}$H+$^4_{\Lambda}$H pairs in such a large amount to be comparable to the production of other twin hypernuclei or double hypernuclei. We therefore conclude, that interactions of quasi-free produced $\Xi^-$ hyperons within the $^9$Be target can not solve the E906 puzzle.

\begin{figure*}[!htb]
\includegraphics[width=0.49\textwidth] {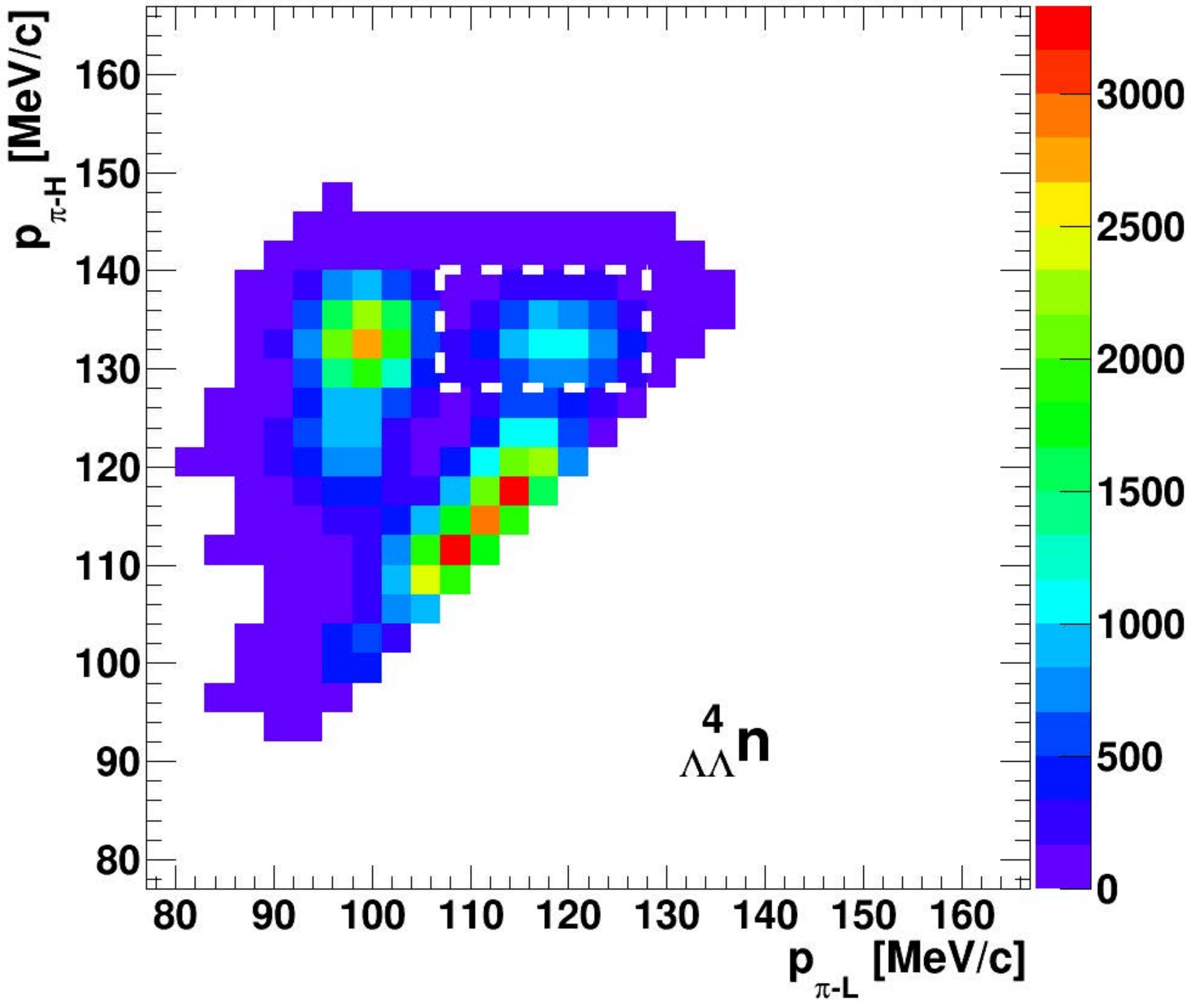}
\includegraphics[width=0.49\textwidth] {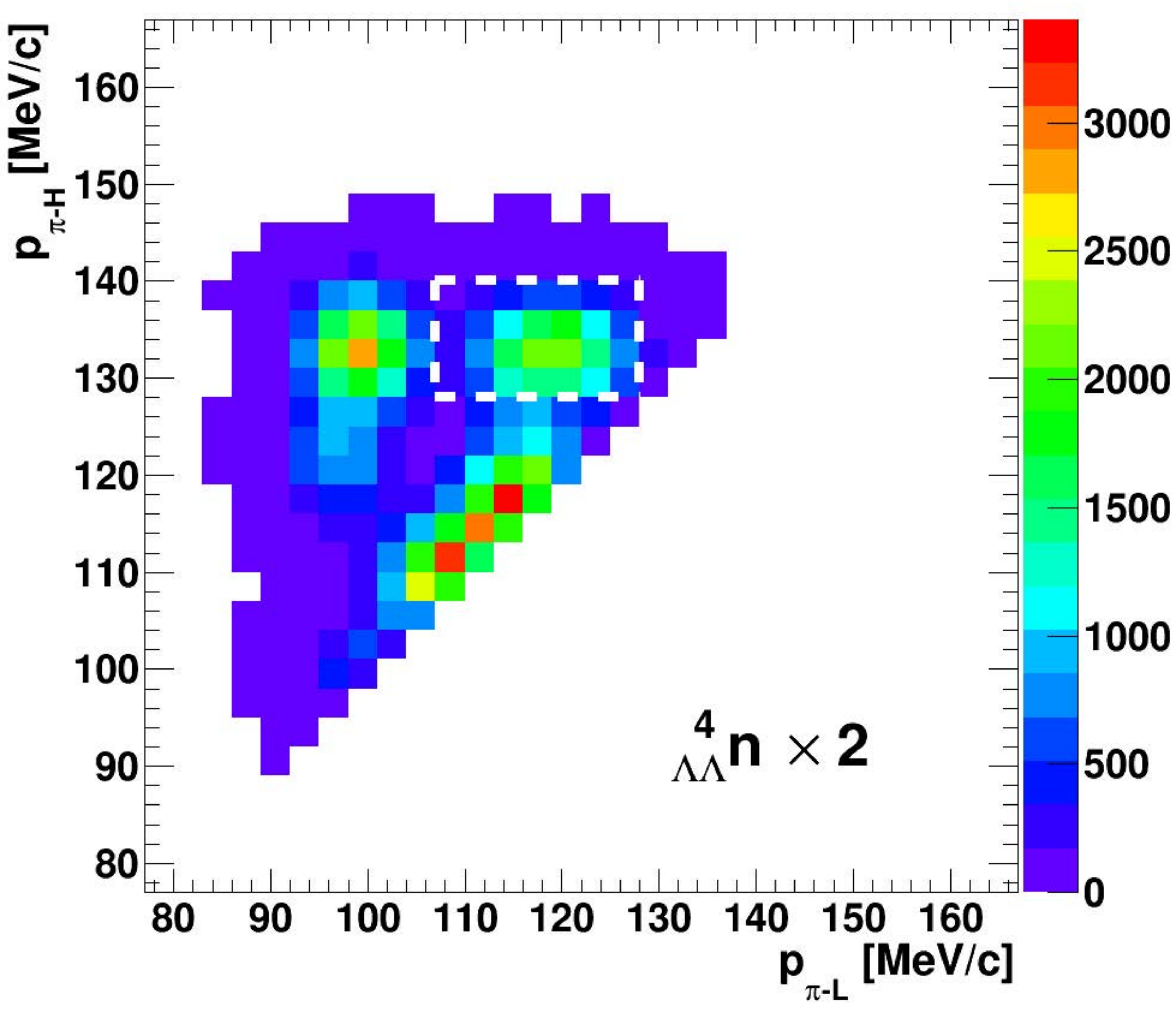}
\includegraphics[width=0.49\textwidth] {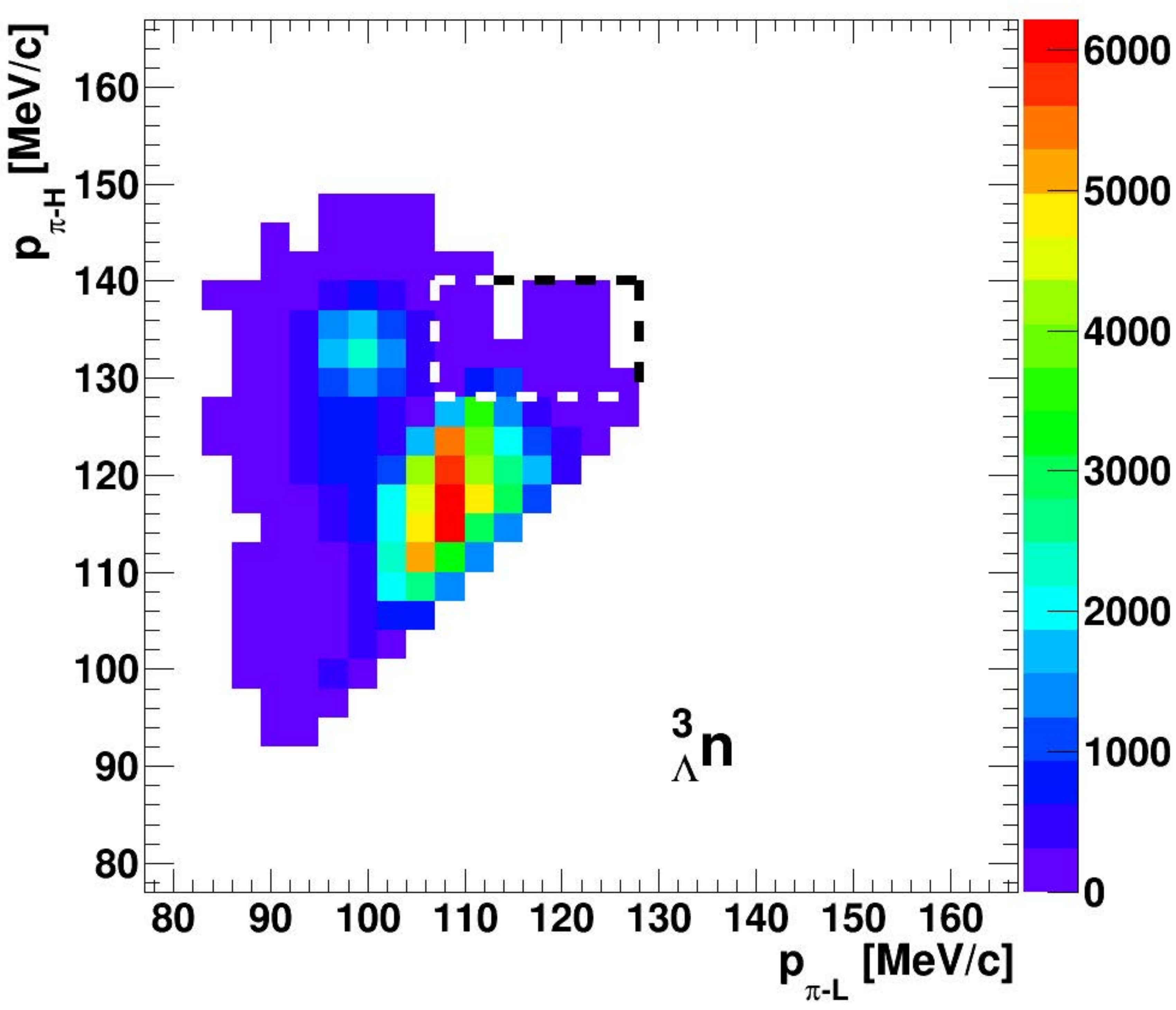}
\includegraphics[width=0.49\textwidth] {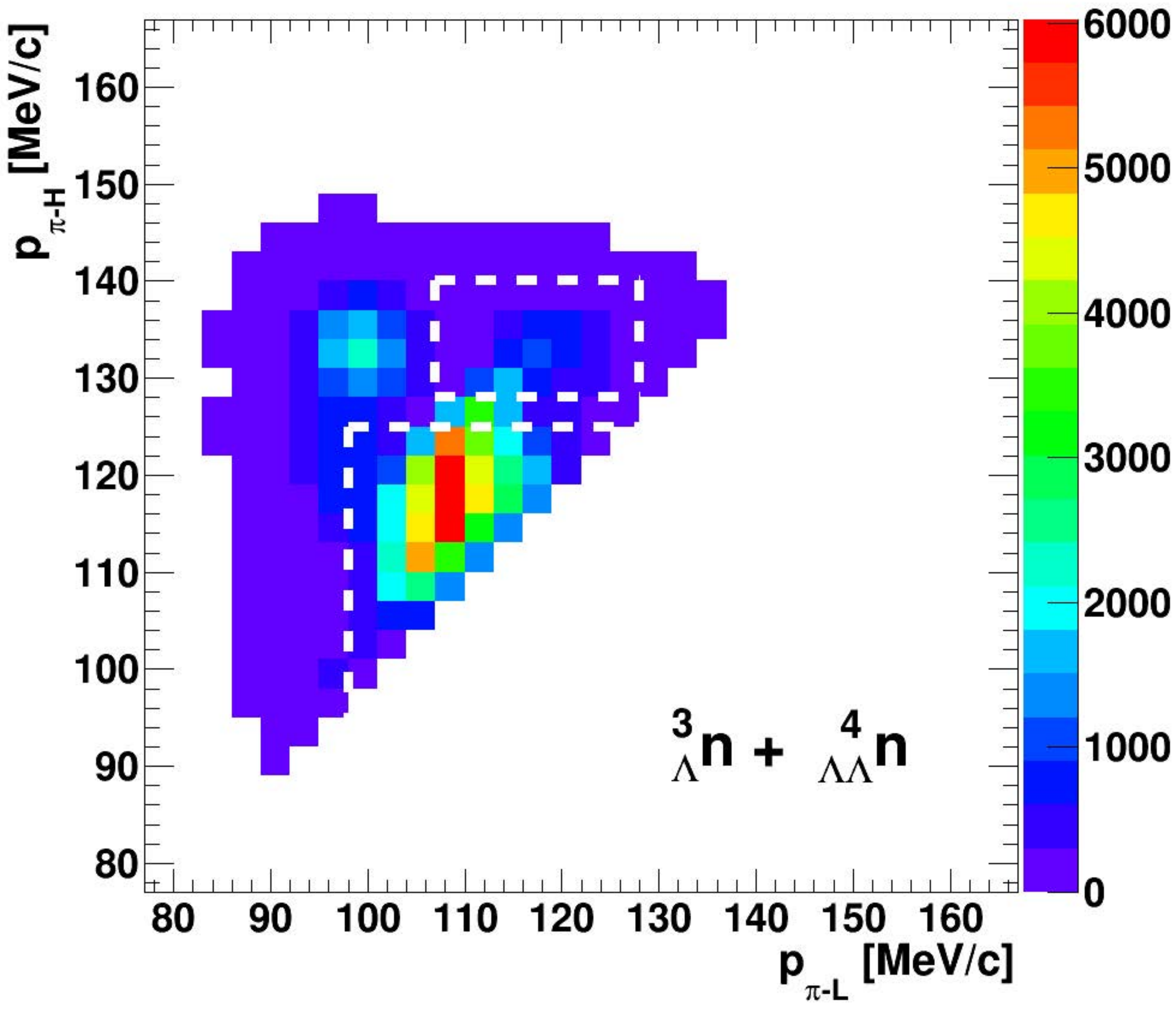}
\caption{Momenta of two correlated pions predicted by the statistical decay model which includes the production and decay of hypothetical neutral \ce{_{$\Lambda\Lambda$}^{4}n} and \ce{_{$\Lambda$}^{3}n} hypernuclei.
In the upper row only \ce{_{$\Lambda\Lambda$}^{4}n} were allowed with a pionic two-body branching ratio of 25\% (left) and 50\% (right), respectively.
The lower, left plot shows the case where only \ce{_{$\Lambda$}^{3}n} is considered. In the lower right panel both neutral hypernuclei were taken into account. The box in each plot marks the region of interest where the enhancement by the E906 experiment has been observed.
}
\label{fig:double:smm_pipi}
\end{figure*}

\section{Can a neutral \ce{_{$\Lambda\Lambda$}^{4}n} nucleus solve the puzzle?}

Motivated by the possible $\pi^-$ momenta (see Tab.~\ref{tab:double:pipi}), Avraham Gal suggested the sequential weak decay of a hypothetical neutral \ce{_{$\Lambda\Lambda$}^{4}n} as a  source for the (133,114)\,MeV/{\em{c}} structure, \cite{Gal2003-private,HYP2003-Pile}. However, this neutral ${\Lambda\Lambda}$-hypernucleus has not been observed yet and its stability is still controversial \cite{PhysRevC.91.014003,PhysRevC.89.061302,1674-1137-41-7-074102}.

In order to test this hypothesis, we implemented the \ce{_{$\Lambda\Lambda$}^{4}n} as a possible decay product in the SMM model. To fix its mass, we assumed a binding energy of B$_{\Lambda\Lambda}$ = 3\,MeV.
In addition, we included the $^{3}_{\Lambda}$n nucleus in some of the calculations. Hints for this neutral hypernucleus have been presented by the HypHI collaboration  \cite{PhysRevC.88.041001}. However, this nucleus is also highly controversial \cite{0305-4616-13-5-002,GAL201493,PhysRevC.89.061302,PhysRevC.89.057001}
and clearly needs further experimental confirmation. The default values for the assumed two-body $\pi^-$ branching ratios and the corresponding $\pi^-$ momenta are listed in Tab.~\ref{tab:double:pipi}.

Besides the experimental momentum resolution of 3.5\,MeV/{\em{c}} of E906, one has to bear in mind that these neutral hypernuclei will not be stopped prior to their weak decay, but instead will decay in flight. In the SMM model, the kinetic energy of the \ce{_{$\Lambda\Lambda$}^{4}n} and $^{3}_{\Lambda}$n
were calculated event-by-event and found to be rather low. For orientation, in case of an initial \ce{_{$\Lambda\Lambda$}^{10}Li^*} at 29.5\,MeV excitation energy, the produced \ce{_{$\Lambda\Lambda$}^{4}n} nuclei have an average kinetic energy of about 2.4\,MeV.
This kinematic smearing causes an additional broadening of the (133,114)\,MeV/{\em{c}} bump by about 9\,MeV/{\em{c}} (FWHM) in horizontal direction which corresponds to 3 bins in Fig.~\ref{fig:double:E906_corr}. Similarly, the neutral $^{3}_{\Lambda}$n single hypernucleus will decay in flight, leading to an additional smearing of the $\pi^-$ momentum from its two-body decay as well.

\begin{figure}
\begin{center}
\includegraphics[width=0.6\columnwidth]{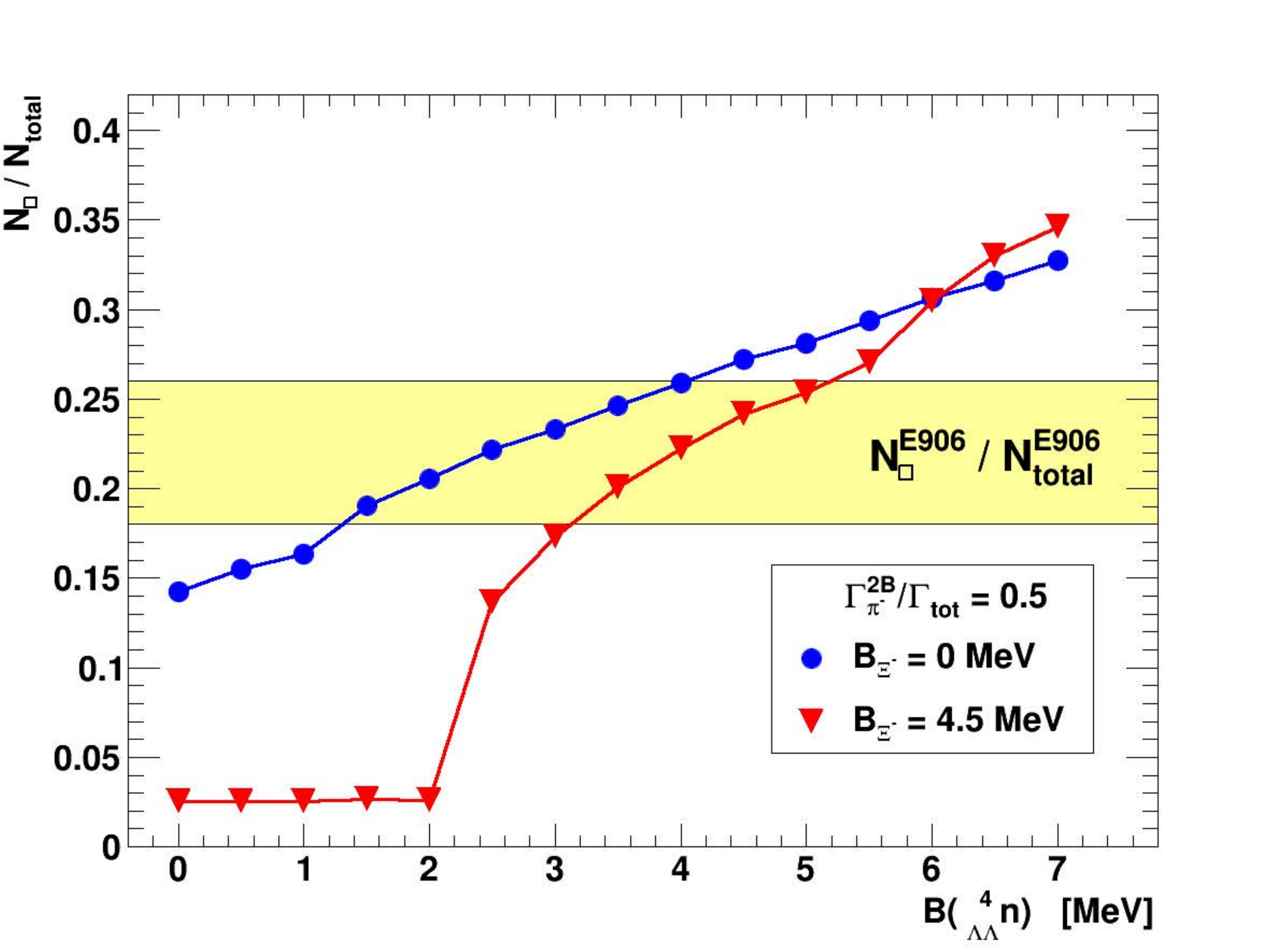}
\end{center}
    \caption{
    Relative strength N$_{\Box}$/N$_{total}$ of the structure at (133,114)\,MeV/{\em{c}} predicted by the statistical multifragmentation model as a function of the
    assumed binding energy of the \ce{_{$\Lambda\Lambda$}^{4}n} nucleus. The two different curves show results for different values for the $\Xi^-$ binding energy at the time of the p$\Xi^-\rightarrow\Lambda\Lambda$ conversion. For the \ce{_{$\Lambda\Lambda$}^{4}n} nucleus a pionic branching ratio of $\Gamma_{\pi^-}^{2B}$/$\Gamma_{tot}$ = 50\% was assumed in all calculations. The $^{3}_{\Lambda}$n nucleus was not allowed as a decay channel.
    }
\label{fig:double:SMM_BXi}
\end{figure}

As before, we consider a $\Xi^-$ capture and conversion process at rest as the principle production scheme, thus forming a \ce{_{$\Lambda\Lambda$}^{10}Li^*} system with 29.5\,MeV excitation energy. Figure \ref{fig:double:smm_pipi} shows the momentum correlations of produced pion pairs for different conditions. These plots should be compared to the right panel of Fig.~\ref{fig:double:E906_corr}, where no neutral hypernuclei were allowed. Numbers for the production probability for the different channels  prior to the weak decay are listed in
Tab.~\ref{tab:double:pipi}. In the upper left panel in Fig.~\ref{fig:double:smm_pipi} only \ce{_{$\Lambda\Lambda$}^{4}n} has been allowed as an additional decay product. For the pionic decay \ce{_{$\Lambda\Lambda$}^{4}n}$\rightarrow \pi^- +^4_{\Lambda}$H a two-body branching ratio of $\Gamma_{\pi^-}^{2B}$/$\Gamma_{tot}$ = 25\% was adopted as the default value. Compared to Fig.~\ref{fig:double:E906_corr}, a clear enhancement in the region of interest marked by the rectangle is visible. Increasing the branching ratio to 50\% (upper, right panel) produces a local structure which resembles the experimental observation of E906 in Fig.~\ref{fig:double:E906_corr}.
Indeed, calculating the ratio between the number of events within the white rectangle and the total number of entries in the upper right plot, we find a value N$_{\Box}$/N$_{total}$~=~0.23. This is remarkable close to the value of 0.22$\pm$0.04 previously determined for the E906 data.

Besides the branching ratio, the binding energy of the \ce{_{$\Lambda\Lambda$}^{4}n} nucleus and the assumed binding energy of the $\Xi^-$ hyperon producing the two $\Lambda$ hyperons determine the strength of the (133,114)\,MeV/{\em{c}} structure in the SMM calculations.
The blue dots in Figure~\ref{fig:double:SMM_BXi} summarize the dependence of the relative strength N$_{\Box}$/N$_{total}$ of the structure at (133,114)\,MeV/{\em{c}} on the assumed binding energy of the \ce{_{$\Lambda\Lambda$}^{4}n} nucleus. A two-body pionic branching ratio of $\Gamma_{\pi^-}^{2B}$/$\Gamma_{tot}$ = 50\% was assumed in all calculations for \ce{_{$\Lambda\Lambda$}^{4}n}. The $^{3}_{\Lambda}$n nucleus was not allowed as a decay channel. The yellow band marks the experimental value of E906. As indicated above, a good agreement is found for \ce{_{$\Lambda\Lambda$}^{4}n} binding energies between about 1.5 and 4\,MeV. Considering possible systematic uncertainties in the E906 data and in the SMM calculations, even
for B(\ce{_{$\Lambda\Lambda$}^{4}n}) = 1\,MeV the signal strength of N$_{\Box}$/N$_{total}$ = 0.16 is still compatible with the data.

As mentioned before, the initial excitation energy of the \ce{_{$\Lambda\Lambda$}^{10}Li^*} depends on the binding energy of the $\Xi^-$ at the time of conversion. We, therefore, repeated the SMM calculations for an excitation energy of 25\,MeV. This is equivalent to a $\Xi^-$ binding energy of 4.5\,MeV, thus representing the capture from a hypernuclear bound state. In this scenario, the SMM calculations shown by the red triangles in Fig.~\ref{fig:double:SMM_BXi} require a \ce{_{$\Lambda\Lambda$}^{4}n} binding energy of at least 3\,MeV to reproduce the E906 data. A smaller two-body pionic branching ratio of \ce{_{$\Lambda\Lambda$}^{4}n} would increase this value even further which would clearly be outside of our present understanding of the \ce{_{$\Lambda\Lambda$}^{4}n} structure.

\section{What about the astounding \ce{_{$\Lambda$}^{3}n} nucleus?}
If the \ce{_{$\Lambda$}^{3}n} nucleus is included as a possible decay channel,
the threshold for $^{3}_{\Lambda}$n+$^7_{\Lambda}$Li production is as low as 21.3\,MeV above the \ce{_{$\Lambda\Lambda$}^{10}Li} ground state. Because of this rather low threshold, the production probability for this decay channel is rather high at the nominal excitation energy of 29.5\,MeV. As a consequence a pronounced structure appears at (108,119)\,MeV/{\em{c}} which is caused by two-body decays of $^{3}_{\Lambda}$n+$^{7}_{\Lambda}$Li twins (see lower parts in Fig.~\ref{fig:double:smm_pipi}). Such a singular enhancement is clearly not consistent with the E906 data. Including additionally the \ce{_{$\Lambda\Lambda$}^{4}n} nucleus (lower, right panel in Fig.~\ref{fig:double:smm_pipi}), does not help to remove this inconsistency.

\begin{figure}
\begin{center}
\includegraphics[width=0.6\columnwidth]{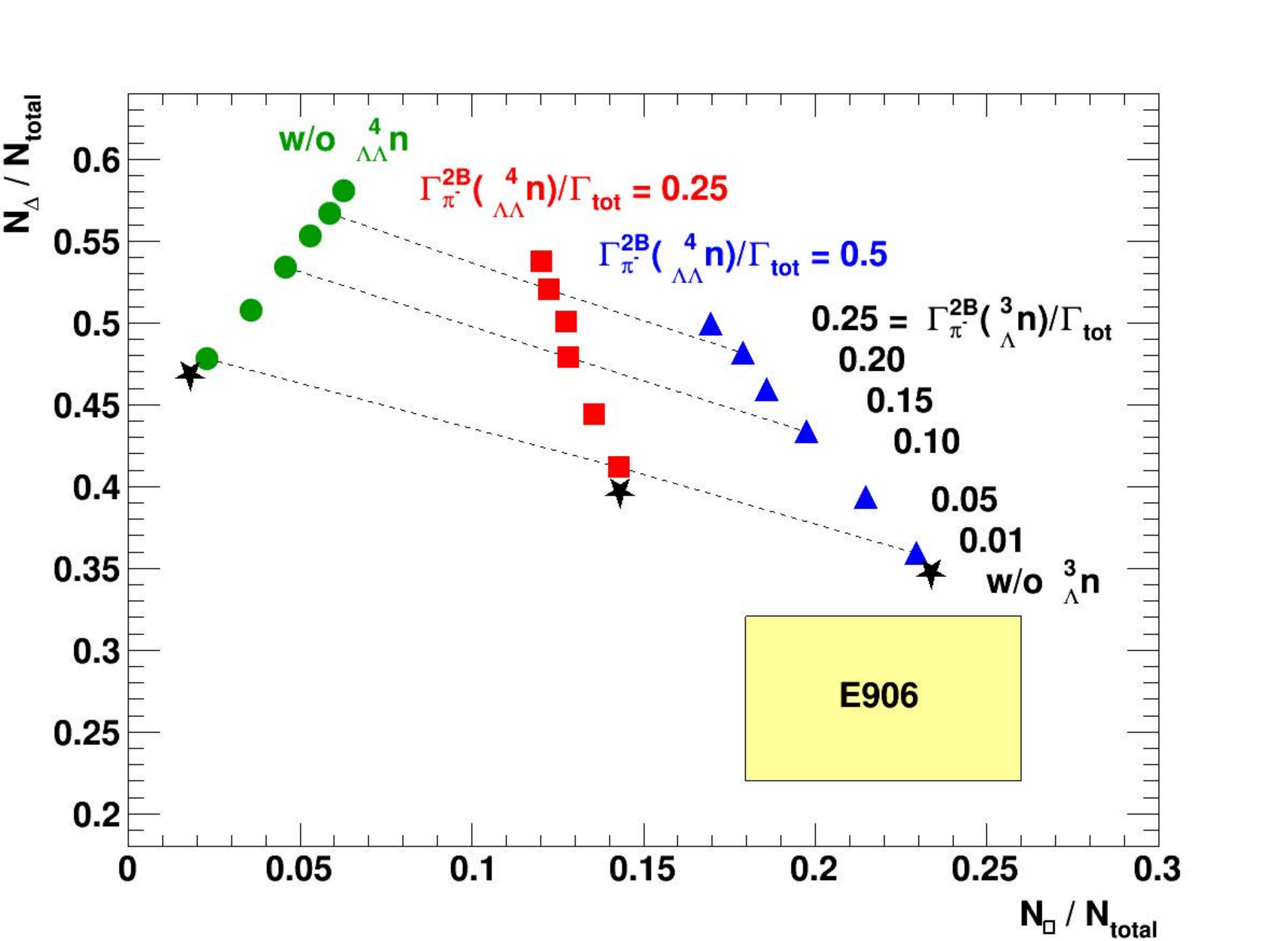}
\end{center}
    \caption{
    Relative strength N$_{\Box}$/N$_{total}$ of the structure at (133,114)\,MeV/{\em{c}} versus the strength N$_{\triangle}$/N$_{total}$ of the enhancement around (108,119)\,MeV/{\em{c}} predicted by the SMM model for different assumptions on the two-body pionic branching ratios of the neutral hypernuclei.
    The rectangular area marks the region suggested by the E906 experiment.
    }
\label{fig:double:SMM_BXi2}
\end{figure}

To study this quantitatively, we define a triangular signal area by p$_{\pi-L} >$ 98\,MeV/{\em{c}} and p$_{\pi-H} <$ 125\,MeV/{\em{c}} (see lower right plot in Fig.~\ref{fig:double:smm_pipi}). Subtracting the  background as before, we obtain an experimental relative signal strength for this region of N$_{\triangle}^{E906}$/N$_{total}^{E906}$ = 0.27 $\pm$ 0.05$_{stat}$. The rectangle in Fig.~\ref{fig:double:SMM_BXi2}
marks the E906 data in the  N$_{\triangle}$/N$_{total}$ vs. N$_{\Box}$/N$_{total}$  plane. The colored symbols are results of SMM calculations. All simulations which discard the production of \ce{_{$\Lambda\Lambda$}^{4}n} (green dots) are far away from the experimental values independent on the two-body branching ratio. Only if \ce{_{$\Lambda\Lambda$}^{4}n} is included {\em and} if the two-body branching ratio for $^{3}_{\Lambda}$n is very small or switched off (black stars), the experimental values of E906 are approached. Finally we note, that including three-body decays of $^6_{\Lambda}$He (see discussion above) would move the lower right black star into the rectangle.

The HypHI measurement also indicates strong contributions from multibody decays $^{3}_{\Lambda}$n$\rightarrow$ d+n+$\pi^-$
\cite{PhysRevC.88.041001}. Because of the low Q-value, pions from this decay will also contribute to the region around (108,119)\,MeV/{\em{c}}. In such a situation, even a very small two-body branching ratio may not be sufficient to avoid the pronounced structure at (108,119)\,MeV/{\em{c}}. Therefore, if the underlying picture of the hypernucleus production is indeed correct, the present SMM calculations make it highly unlikely that a bound $^{3}_{\Lambda}$n nucleus is produced in the E906 experiment.

Before concluding this section, we return to the \ce{_{$\Lambda\Lambda$}^{4}H} nucleus, which has originally been suggested as the main source of the (114,104)\,MeV/{\em{c}} structure.
Compared to the SMM calculations for stopped $\Xi^-$ hyperons,
the GiBUU simulations for $\Xi^-$ + $^9$Be reactions predict an additional contribution to the production of
\ce{_{$\Lambda\Lambda$}^{4}H} which is about a factor of 2 larger. Still, its summed production probability is a factor of 4 lower than the one for \ce{_{$\Lambda\Lambda$}^{7}He}. Thus, \ce{_{$\Lambda\Lambda$}^{4}H} decays may indeed contribute to the (114,104)\,MeV/{\em{c}} structure. However, the large momentum broadening of about 10\,MeV/{\em{c}} (FWHM), caused by  \ce{_{$\Lambda$}^{4}He} $\rightarrow$ $^4$Li decays will prevent the formation of a prominent structure in the $\pi^-$-$\pi^-$ momentum correlation.

\section{Conclusion}
\label{sec:hyp_decay:conlusion}
Neutral nuclei remain a challenge for the field of conventional nuclei as well as for hypernuclear studies. Despite many recent promising experimental results, the question of their stability is not yet answered beyond doubt. The E906 experiment was the first fully electronic experiment to produce and study double hypernuclei with large statistics. Unfortunately, the interpretation of the measured $\pi^-$--$\pi^-$ momentum correlation is still blurry
because the hypothesized production of $^3_{\Lambda}$H+$^4_{\Lambda}$H pairs remains questionable. In this paper we have shown that neither a scenario where the double hypernuclei are produced after capture of a stopped $\Xi^-$ nor reactions of energetic $\Xi^-$ with $^9$Be nuclei in the target can produce a sufficient amount of such pairs.
We have therefore explored the conjecture that decays of the \ce{_{$\Lambda\Lambda$}^{4}n} may be responsible for the observed structure. Considering bound \ce{_{$\Lambda\Lambda$}^{4}n} production with a two-body $\pi^-$ branching ratio of 50\% in the statistical multifragmentation model, this describes the measured data of the E906 experiment remarkably well.
At the same time these calculations exclude the existence of a bound $^{3}_{\Lambda}$n with a large two-body $\pi^-$ branching ratio in the range of 25\%.

Clearly, this study should not be considered as a direct proof for the existence of a bound neutral double hypernucleus \ce{_{$\Lambda\Lambda$}^{4}n}. Nonetheless, the proposed scenario
provides at present the most consistent explanation for the pion momentum correlations observed by the E906 experiment. Experimentally, a search for a direct detection of \ce{_{$\Lambda\Lambda$}^{4}n} could e.g. be performed by the current J-PARC E07 emulsion experiment. In particular, the newly developed so called 'vertex picker' scanning system, which does not rely on the incoming $\Xi^-$ track \cite{YOSHIDA201786} could be very efficient to identify isolated two-vertex events from in-flight decays of \ce{_{$\Lambda\Lambda$}^{4}n}. In that sense, we hope that this work will stimulate further discussions and experimental activities which  -- finally -- may help to solve the E906 puzzle.

\section*{Acknowledgements}
We thank Alexander Botvina for providing us with the original statistical multifragmentation model code.
This work uses data collected within the framework of the PhD thesis of S. Bleser
at the Johannes Gutenberg University Mainz. Work supported in part by COST-Actions (THOR) CA15213 and (PHAROS) CA16214.


\begin{thebibliography}{65}%
\makeatletter
\providecommand \@ifxundefined [1]{%
 \@ifx{#1\undefined}
}%
\providecommand \@ifnum [1]{%
 \ifnum #1\expandafter \@firstoftwo
 \else \expandafter \@secondoftwo
 \fi
}%
\providecommand \@ifx [1]{%
 \ifx #1\expandafter \@firstoftwo
 \else \expandafter \@secondoftwo
 \fi
}%
\providecommand \natexlab [1]{#1}%
\providecommand \enquote  [1]{``#1''}%
\providecommand \bibnamefont  [1]{#1}%
\providecommand \bibfnamefont [1]{#1}%
\providecommand \citenamefont [1]{#1}%
\providecommand \href@noop [0]{\@secondoftwo}%
\providecommand \href [0]{\begingroup \@sanitize@url \@href}%
\providecommand \@href[1]{\@@startlink{#1}\@@href}%
\providecommand \@@href[1]{\endgroup#1\@@endlink}%
\providecommand \@sanitize@url [0]{\catcode `\$12\catcode `\&12\catcode
  `\#12\catcode `\^12\catcode `\_12\catcode `\%12\relax}%
\providecommand \@@startlink[1]{}%
\providecommand \@@endlink[0]{}%
\providecommand \url  [0]{\begingroup\@sanitize@url \@url }%
\providecommand \@url [1]{\endgroup\@href {#1}{\urlprefix }}%
\providecommand \urlprefix  [0]{URL }%
\providecommand \Eprint [0]{\href }%
\providecommand \doibase [0]{http://dx.doi.org/}%
\providecommand \selectlanguage [0]{\@gobble}%
\providecommand \bibinfo  [0]{\@secondoftwo}%
\providecommand \bibfield  [0]{\@secondoftwo}%
\providecommand \translation [1]{[#1]}%
\providecommand \BibitemOpen [0]{}%
\providecommand \bibitemStop [0]{}%
\providecommand \bibitemNoStop [0]{.\EOS\space}%
\providecommand \EOS [0]{\spacefactor3000\relax}%
\providecommand \BibitemShut  [1]{\csname bibitem#1\endcsname}%
\let\auto@bib@innerbib\@empty
\bibitem [{\citenamefont {Kisamori}\ \emph {et~al.}(2016)\citenamefont
  {Kisamori} \emph {et~al.}}]{PhysRevLett.116.052501}%
  \BibitemOpen
  \bibfield  {author} {\bibinfo {author} {\bibfnamefont {K.}~\bibnamefont
  {Kisamori}} \emph {et~al.},\ }\href {\doibase 10.1103/PhysRevLett.116.052501}
  {\bibfield  {journal} {\bibinfo  {journal} {Phys. Rev. Lett.}\ }\textbf
  {\bibinfo {volume} {116}},\ p.\ \bibinfo {pages} {052501} (\bibinfo {year}
  {2016})}\BibitemShut {NoStop}%
\bibitem [{\citenamefont {Marqu\'{e}s}\ \emph {et~al.}(2002)\citenamefont
  {Marqu\'{e}s} \emph {et~al.}}]{PhysRevC.65.044006}%
  \BibitemOpen
  \bibfield  {author} {\bibinfo {author} {\bibfnamefont {F.~M.}\ \bibnamefont
  {Marqu\'{e}s}} \emph {et~al.},\ }\href {\doibase 10.1103/PhysRevC.65.044006}
  {\bibfield  {journal} {\bibinfo  {journal} {Phys. Rev. C}\ }\textbf {\bibinfo
  {volume} {65}},\ p.\ \bibinfo {pages} {044006} (\bibinfo {year}
  {2002})}\BibitemShut {NoStop}%
\bibitem [{\citenamefont {Rappold}\ \emph {et~al.}(2013)\citenamefont {Rappold}
  \emph {et~al.}}]{PhysRevC.88.041001}%
  \BibitemOpen
  \bibfield  {author} {\bibinfo {author} {\bibfnamefont {C.}~\bibnamefont
  {Rappold}} \emph {et~al.} (\bibinfo {collaboration} {HypHI Collaboration}),\
  }\href {\doibase 10.1103/PhysRevC.88.041001} {\bibfield  {journal} {\bibinfo
  {journal} {Phys. Rev. C}\ }\textbf {\bibinfo {volume} {88}},\ p.\ \bibinfo
  {pages} {041001} (\bibinfo {year} {2013})}\BibitemShut {NoStop}%
\bibitem [{\citenamefont {Pieper}(2003)}]{PhysRevLett.90.252501}%
  \BibitemOpen
  \bibfield  {author} {\bibinfo {author} {\bibfnamefont {S.~C.}\ \bibnamefont
  {Pieper}},\ }\href {\doibase 10.1103/PhysRevLett.90.252501} {\bibfield
  {journal} {\bibinfo  {journal} {Phys. Rev. Lett.}\ }\textbf {\bibinfo
  {volume} {90}},\ p.\ \bibinfo {pages} {252501} (\bibinfo {year}
  {2003})}\BibitemShut {NoStop}%
\bibitem [{\citenamefont {Timofeyuk}(2003)}]{0954-3899-29-2-102}%
  \BibitemOpen
  \bibfield  {author} {\bibinfo {author} {\bibfnamefont {N.}~\bibnamefont
  {Timofeyuk}},\ }\href {http://stacks.iop.org/0954-3899/29/i=2/a=102}
  {\bibfield  {journal} {\bibinfo  {journal} {Journal of Physics G: Nuclear and
  Particle Physics}\ }\textbf {\bibinfo {volume} {29}},\ p.~\bibinfo {pages}
  {L9} (\bibinfo {year} {2003})}\BibitemShut {NoStop}%
\bibitem [{\citenamefont {Bertulani}\ and\ \citenamefont
  {Zelevinsky}(2003)}]{0954-3899-29-10-309}%
  \BibitemOpen
  \bibfield  {author} {\bibinfo {author} {\bibfnamefont {C.}~\bibnamefont
  {Bertulani}}\ and\ \bibinfo {author} {\bibfnamefont {V.}~\bibnamefont
  {Zelevinsky}},\ }\href {http://stacks.iop.org/0954-3899/29/i=10/a=309}
  {\bibfield  {journal} {\bibinfo  {journal} {Journal of Physics G: Nuclear and
  Particle Physics}\ }\textbf {\bibinfo {volume} {29}},\ p.\ \bibinfo {pages}
  {2431} (\bibinfo {year} {2003})}\BibitemShut {NoStop}%
\bibitem [{\citenamefont {Hiyama}\ \emph {et~al.}(2016)\citenamefont {Hiyama},
  \citenamefont {Lazauskas}, \citenamefont {Carbonell},\ and\ \citenamefont
  {Kamimura}}]{PhysRevC.93.044004}%
  \BibitemOpen
  \bibfield  {author} {\bibinfo {author} {\bibfnamefont {E.}~\bibnamefont
  {Hiyama}}, \bibinfo {author} {\bibfnamefont {R.}~\bibnamefont {Lazauskas}},
  \bibinfo {author} {\bibfnamefont {J.}~\bibnamefont {Carbonell}}, \ and\
  \bibinfo {author} {\bibfnamefont {M.}~\bibnamefont {Kamimura}},\ }\href
  {\doibase 10.1103/PhysRevC.93.044004} {\bibfield  {journal} {\bibinfo
  {journal} {Phys. Rev. C}\ }\textbf {\bibinfo {volume} {93}},\ p.\ \bibinfo
  {pages} {044004} (\bibinfo {year} {2016})}\BibitemShut {NoStop}%
\bibitem [{\citenamefont {Fossez}\ \emph {et~al.}(2017)\citenamefont {Fossez},
  \citenamefont {Rotureau}, \citenamefont {Michel},\ and\ \citenamefont
  {P\l{}oszajczak}}]{PhysRevLett.119.032501}%
  \BibitemOpen
  \bibfield  {author} {\bibinfo {author} {\bibfnamefont {K.}~\bibnamefont
  {Fossez}}, \bibinfo {author} {\bibfnamefont {J.}~\bibnamefont {Rotureau}},
  \bibinfo {author} {\bibfnamefont {N.}~\bibnamefont {Michel}}, \ and\ \bibinfo
  {author} {\bibfnamefont {M.}~\bibnamefont {P\l{}oszajczak}},\ }\href
  {\doibase 10.1103/PhysRevLett.119.032501} {\bibfield  {journal} {\bibinfo
  {journal} {Phys. Rev. Lett.}\ }\textbf {\bibinfo {volume} {119}},\ p.\
  \bibinfo {pages} {032501} (\bibinfo {year} {2017})}\BibitemShut {NoStop}%
\bibitem [{\citenamefont {Lashko}\ and\ \citenamefont
  {Filippov}(2008)}]{Lashko2008}%
  \BibitemOpen
  \bibfield  {author} {\bibinfo {author} {\bibfnamefont {Y.}~\bibnamefont
  {Lashko}}\ and\ \bibinfo {author} {\bibfnamefont {G.}~\bibnamefont
  {Filippov}},\ }\href {\doibase 10.1134/S1063778808020014} {\bibfield
  {journal} {\bibinfo  {journal} {Physics of Atomic Nuclei}\ }\textbf {\bibinfo
  {volume} {71}},\ \unskip\ \bibinfo {pages} {209--214} (\bibinfo {year}
  {2008})}\BibitemShut {NoStop}%
\bibitem [{\citenamefont {Shirokov}\ \emph {et~al.}(2016)\citenamefont
  {Shirokov}, \citenamefont {Papadimitriou}, \citenamefont {Mazur},
  \citenamefont {Mazur}, \citenamefont {Roth},\ and\ \citenamefont
  {Vary}}]{PhysRevLett.117.182502}%
  \BibitemOpen
  \bibfield  {author} {\bibinfo {author} {\bibfnamefont {A.}~\bibnamefont
  {Shirokov}}, \bibinfo {author} {\bibfnamefont {G.}~\bibnamefont
  {Papadimitriou}}, \bibinfo {author} {\bibfnamefont {A.}~\bibnamefont
  {Mazur}}, \bibinfo {author} {\bibfnamefont {I.}~\bibnamefont {Mazur}},
  \bibinfo {author} {\bibfnamefont {R.}~\bibnamefont {Roth}}, \ and\ \bibinfo
  {author} {\bibfnamefont {J.}~\bibnamefont {Vary}},\ }\href {\doibase
  10.1103/PhysRevLett.117.182502} {\bibfield  {journal} {\bibinfo  {journal}
  {Phys. Rev. Lett.}\ }\textbf {\bibinfo {volume} {117}},\ p.\ \bibinfo {pages}
  {182502} (\bibinfo {year} {2016})}\BibitemShut {NoStop}%
\bibitem [{\citenamefont {Shirokov}\ \emph {et~al.}(2018)\citenamefont
  {Shirokov}, \citenamefont {Papadimitriou}, \citenamefont {Mazur},
  \citenamefont {Mazur}, \citenamefont {Roth},\ and\ \citenamefont
  {Vary}}]{PhysRevLett.121.099901}%
  \BibitemOpen
  \bibfield  {author} {\bibinfo {author} {\bibfnamefont {A.~M.}\ \bibnamefont
  {Shirokov}}, \bibinfo {author} {\bibfnamefont {G.}~\bibnamefont
  {Papadimitriou}}, \bibinfo {author} {\bibfnamefont {A.~I.}\ \bibnamefont
  {Mazur}}, \bibinfo {author} {\bibfnamefont {I.~A.}\ \bibnamefont {Mazur}},
  \bibinfo {author} {\bibfnamefont {R.}~\bibnamefont {Roth}}, \ and\ \bibinfo
  {author} {\bibfnamefont {J.~P.}\ \bibnamefont {Vary}},\ }\href {\doibase
  10.1103/PhysRevLett.121.099901} {\bibfield  {journal} {\bibinfo  {journal}
  {Phys. Rev. Lett.}\ }\textbf {\bibinfo {volume} {121}},\ p.\ \bibinfo {pages}
  {099901} (\bibinfo {year} {2018})}\BibitemShut {NoStop}%
\bibitem [{\citenamefont {Gandolfi}\ \emph {et~al.}(2017)\citenamefont
  {Gandolfi}, \citenamefont {Hammer}, \citenamefont {Klos}, \citenamefont
  {Lynn},\ and\ \citenamefont {Schwenk}}]{PhysRevLett.118.232501}%
  \BibitemOpen
  \bibfield  {author} {\bibinfo {author} {\bibfnamefont {S.}~\bibnamefont
  {Gandolfi}}, \bibinfo {author} {\bibfnamefont {H.-W.}\ \bibnamefont
  {Hammer}}, \bibinfo {author} {\bibfnamefont {P.}~\bibnamefont {Klos}},
  \bibinfo {author} {\bibfnamefont {J.~E.}\ \bibnamefont {Lynn}}, \ and\
  \bibinfo {author} {\bibfnamefont {A.}~\bibnamefont {Schwenk}},\ }\href
  {\doibase 10.1103/PhysRevLett.118.232501} {\bibfield  {journal} {\bibinfo
  {journal} {Phys. Rev. Lett.}\ }\textbf {\bibinfo {volume} {118}},\ p.\
  \bibinfo {pages} {232501} (\bibinfo {year} {2017})}\BibitemShut {NoStop}%
\bibitem [{\citenamefont {Garcilazo}(1987)}]{0305-4616-13-5-002}%
  \BibitemOpen
  \bibfield  {author} {\bibinfo {author} {\bibfnamefont {H.}~\bibnamefont
  {Garcilazo}},\ }\href {http://stacks.iop.org/0305-4616/13/i=5/a=002}
  {\bibfield  {journal} {\bibinfo  {journal} {Journal of Physics G: Nuclear
  Physics}\ }\textbf {\bibinfo {volume} {13}},\ p.\ \bibinfo {pages} {L63}
  (\bibinfo {year} {1987})}\BibitemShut {NoStop}%
\bibitem [{\citenamefont {Gal}\ and\ \citenamefont
  {Garcilazo}(2014)}]{GAL201493}%
  \BibitemOpen
  \bibfield  {author} {\bibinfo {author} {\bibfnamefont {A.}~\bibnamefont
  {Gal}}\ and\ \bibinfo {author} {\bibfnamefont {H.}~\bibnamefont
  {Garcilazo}},\ }\href {\doibase
  https://doi.org/10.1016/j.physletb.2014.07.009} {\bibfield  {journal}
  {\bibinfo  {journal} {Physics Letters B}\ }\textbf {\bibinfo {volume}
  {736}},\ \unskip\ \bibinfo {pages} {93 -- 97} (\bibinfo {year}
  {2014})}\BibitemShut {NoStop}%
\bibitem [{\citenamefont {Hiyama}\ \emph {et~al.}(2014)\citenamefont {Hiyama},
  \citenamefont {Ohnishi}, \citenamefont {Gibson},\ and\ \citenamefont
  {Rijken}}]{PhysRevC.89.061302}%
  \BibitemOpen
  \bibfield  {author} {\bibinfo {author} {\bibfnamefont {E.}~\bibnamefont
  {Hiyama}}, \bibinfo {author} {\bibfnamefont {S.}~\bibnamefont {Ohnishi}},
  \bibinfo {author} {\bibfnamefont {B.~F.}\ \bibnamefont {Gibson}}, \ and\
  \bibinfo {author} {\bibfnamefont {T.~A.}\ \bibnamefont {Rijken}},\ }\href
  {\doibase 10.1103/PhysRevC.89.061302} {\bibfield  {journal} {\bibinfo
  {journal} {Phys. Rev. C}\ }\textbf {\bibinfo {volume} {89}},\ p.\ \bibinfo
  {pages} {061302} (\bibinfo {year} {2014})}\BibitemShut {NoStop}%
\bibitem [{\citenamefont {Garcilazo}\ and\ \citenamefont
  {Valcarce}(2014)}]{PhysRevC.89.057001}%
  \BibitemOpen
  \bibfield  {author} {\bibinfo {author} {\bibfnamefont {H.}~\bibnamefont
  {Garcilazo}}\ and\ \bibinfo {author} {\bibfnamefont {A.}~\bibnamefont
  {Valcarce}},\ }\href {\doibase 10.1103/PhysRevC.89.057001} {\bibfield
  {journal} {\bibinfo  {journal} {Phys. Rev. C}\ }\textbf {\bibinfo {volume}
  {89}},\ p.\ \bibinfo {pages} {057001} (\bibinfo {year} {2014})}\BibitemShut
  {NoStop}%
\bibitem [{\citenamefont {Garcilazo}, \citenamefont {Valcarce},\ and\
  \citenamefont {Vijande}(2017)}]{1674-1137-41-7-074102}%
  \BibitemOpen
  \bibfield  {author} {\bibinfo {author} {\bibfnamefont {H.}~\bibnamefont
  {Garcilazo}}, \bibinfo {author} {\bibfnamefont {A.}~\bibnamefont {Valcarce}},
  \ and\ \bibinfo {author} {\bibfnamefont {J.}~\bibnamefont {Vijande}},\ }\href
  {http://stacks.iop.org/1674-1137/41/i=7/a=074102} {\bibfield  {journal}
  {\bibinfo  {journal} {Chinese Physics C}\ }\textbf {\bibinfo {volume} {41}},\
  p.\ \bibinfo {pages} {074102} (\bibinfo {year} {2017})}\BibitemShut {NoStop}%
\bibitem [{\citenamefont {Nemura}, \citenamefont {Akaishi},\ and\ \citenamefont
  {Suzuki}(2002)}]{PhysRevLett.89.142504}%
  \BibitemOpen
  \bibfield  {author} {\bibinfo {author} {\bibfnamefont {H.}~\bibnamefont
  {Nemura}}, \bibinfo {author} {\bibfnamefont {Y.}~\bibnamefont {Akaishi}}, \
  and\ \bibinfo {author} {\bibfnamefont {Y.}~\bibnamefont {Suzuki}},\ }\href
  {\doibase 10.1103/PhysRevLett.89.142504} {\bibfield  {journal} {\bibinfo
  {journal} {Phys. Rev. Lett.}\ }\textbf {\bibinfo {volume} {89}},\ p.\
  \bibinfo {pages} {142504}Sep (\bibinfo {year} {2002})}\BibitemShut {NoStop}%
\bibitem [{\citenamefont {Filikhin}\ and\ \citenamefont
  {Gal}(2002{\natexlab{a}})}]{PhysRevLett.89.172502}%
  \BibitemOpen
  \bibfield  {author} {\bibinfo {author} {\bibfnamefont {I.~N.}\ \bibnamefont
  {Filikhin}}\ and\ \bibinfo {author} {\bibfnamefont {A.}~\bibnamefont {Gal}},\
  }\href {\doibase 10.1103/PhysRevLett.89.172502} {\bibfield  {journal}
  {\bibinfo  {journal} {Phys. Rev. Lett.}\ }\textbf {\bibinfo {volume} {89}},\
  p.\ \bibinfo {pages} {172502} (\bibinfo {year}
  {2002}{\natexlab{a}})}\BibitemShut {NoStop}%
\bibitem [{\citenamefont {Nemura}, \citenamefont {Akaishi},\ and\ \citenamefont
  {Myint}(2003)}]{PhysRevC.67.051001}%
  \BibitemOpen
  \bibfield  {author} {\bibinfo {author} {\bibfnamefont {H.}~\bibnamefont
  {Nemura}}, \bibinfo {author} {\bibfnamefont {Y.}~\bibnamefont {Akaishi}}, \
  and\ \bibinfo {author} {\bibfnamefont {K.~S.}\ \bibnamefont {Myint}},\ }\href
  {\doibase 10.1103/PhysRevC.67.051001} {\bibfield  {journal} {\bibinfo
  {journal} {Phys. Rev. C}\ }\textbf {\bibinfo {volume} {67}},\ p.\ \bibinfo
  {pages} {051001} (\bibinfo {year} {2003})}\BibitemShut {NoStop}%
\bibitem [{\citenamefont {Nemura}\ \emph {et~al.}(2005)\citenamefont {Nemura},
  \citenamefont {Shinmura}, \citenamefont {Akaishi},\ and\ \citenamefont
  {Myint}}]{PhysRevLett.94.202502}%
  \BibitemOpen
  \bibfield  {author} {\bibinfo {author} {\bibfnamefont {H.}~\bibnamefont
  {Nemura}}, \bibinfo {author} {\bibfnamefont {S.}~\bibnamefont {Shinmura}},
  \bibinfo {author} {\bibfnamefont {Y.}~\bibnamefont {Akaishi}}, \ and\
  \bibinfo {author} {\bibfnamefont {K.~S.}\ \bibnamefont {Myint}},\ }\href
  {\doibase 10.1103/PhysRevLett.94.202502} {\bibfield  {journal} {\bibinfo
  {journal} {Phys. Rev. Lett.}\ }\textbf {\bibinfo {volume} {94}},\ p.\
  \bibinfo {pages} {202502} (\bibinfo {year} {2005})}\BibitemShut {NoStop}%
\bibitem [{\citenamefont {Filikhin}\ and\ \citenamefont
  {Gal}(2002{\natexlab{b}})}]{FILIKHIN2002491}%
  \BibitemOpen
  \bibfield  {author} {\bibinfo {author} {\bibfnamefont {I.}~\bibnamefont
  {Filikhin}}\ and\ \bibinfo {author} {\bibfnamefont {A.}~\bibnamefont {Gal}},\
  }\href {\doibase http://dx.doi.org/10.1016/S0375-9474(02)01008-4} {\bibfield
  {journal} {\bibinfo  {journal} {Nuclear Physics A}\ }\textbf {\bibinfo
  {volume} {707}},\ \unskip\ \bibinfo {pages} {491 -- 509} (\bibinfo {year}
  {2002}{\natexlab{b}})}\BibitemShut {NoStop}%
\bibitem [{\citenamefont {Swe~Myint}, \citenamefont {Shinmura},\ and\
  \citenamefont {Akaishi}(2003)}]{SweMyint2003}%
  \BibitemOpen
  \bibfield  {author} {\bibinfo {author} {\bibfnamefont {K.}~\bibnamefont
  {Swe~Myint}}, \bibinfo {author} {\bibfnamefont {S.}~\bibnamefont {Shinmura}},
  \ and\ \bibinfo {author} {\bibfnamefont {Y.}~\bibnamefont {Akaishi}},\ }\href
  {\doibase 10.1140/epja/i2002-10083-y} {\bibfield  {journal} {\bibinfo
  {journal} {The European Physical Journal A - Hadrons and Nuclei}\ }\textbf
  {\bibinfo {volume} {16}},\ \unskip\ \bibinfo {pages} {21--26} (\bibinfo
  {year} {2003})}\BibitemShut {NoStop}%
\bibitem [{\citenamefont {Gal}(2005)}]{GAL200591}%
  \BibitemOpen
  \bibfield  {author} {\bibinfo {author} {\bibfnamefont {A.}~\bibnamefont
  {Gal}},\ }\href {\doibase http://dx.doi.org/10.1016/j.nuclphysa.2005.01.021}
  {\bibfield  {journal} {\bibinfo  {journal} {Nuclear Physics A}\ }\textbf
  {\bibinfo {volume} {754}},\ \unskip\ \bibinfo {pages} {91 -- 102} (\bibinfo
  {year} {2005})}\BibitemShut {NoStop}%
\bibitem [{\citenamefont {Garcilazo}\ and\ \citenamefont
  {Valcarce}(2013{\natexlab{a}})}]{PhysRevLett.110.012503}%
  \BibitemOpen
  \bibfield  {author} {\bibinfo {author} {\bibfnamefont {H.}~\bibnamefont
  {Garcilazo}}\ and\ \bibinfo {author} {\bibfnamefont {A.}~\bibnamefont
  {Valcarce}},\ }\href {\doibase 10.1103/PhysRevLett.110.012503} {\bibfield
  {journal} {\bibinfo  {journal} {Phys. Rev. Lett.}\ }\textbf {\bibinfo
  {volume} {110}},\ p.\ \bibinfo {pages} {012503} (\bibinfo {year}
  {2013}{\natexlab{a}})}\BibitemShut {NoStop}%
\bibitem [{\citenamefont {Gal}(2013)}]{PhysRevLett.110.179201}%
  \BibitemOpen
  \bibfield  {author} {\bibinfo {author} {\bibfnamefont {A.}~\bibnamefont
  {Gal}},\ }\href {\doibase 10.1103/PhysRevLett.110.179201} {\bibfield
  {journal} {\bibinfo  {journal} {Phys. Rev. Lett.}\ }\textbf {\bibinfo
  {volume} {110}},\ p.\ \bibinfo {pages} {179201} (\bibinfo {year}
  {2013})}\BibitemShut {NoStop}%
\bibitem [{\citenamefont {Garcilazo}\ and\ \citenamefont
  {Valcarce}(2013{\natexlab{b}})}]{PhysRevLett.110.179202}%
  \BibitemOpen
  \bibfield  {author} {\bibinfo {author} {\bibfnamefont {H.}~\bibnamefont
  {Garcilazo}}\ and\ \bibinfo {author} {\bibfnamefont {A.}~\bibnamefont
  {Valcarce}},\ }\href {\doibase 10.1103/PhysRevLett.110.179202} {\bibfield
  {journal} {\bibinfo  {journal} {Phys. Rev. Lett.}\ }\textbf {\bibinfo
  {volume} {110}},\ p.\ \bibinfo {pages} {179202} (\bibinfo {year}
  {2013}{\natexlab{b}})}\BibitemShut {NoStop}%
\bibitem [{\citenamefont {Richard}, \citenamefont {Wang},\ and\ \citenamefont
  {Zhao}(2014)}]{1742-6596-569-1-012079}%
  \BibitemOpen
  \bibfield  {author} {\bibinfo {author} {\bibfnamefont {J.-M.}\ \bibnamefont
  {Richard}}, \bibinfo {author} {\bibfnamefont {Q.}~\bibnamefont {Wang}}, \
  and\ \bibinfo {author} {\bibfnamefont {Q.}~\bibnamefont {Zhao}},\ }\href
  {http://stacks.iop.org/1742-6596/569/i=1/a=012079} {\bibfield  {journal}
  {\bibinfo  {journal} {Journal of Physics: Conference Series}\ }\textbf
  {\bibinfo {volume} {569}},\ p.\ \bibinfo {pages} {012079} (\bibinfo {year}
  {2014})}\BibitemShut {NoStop}%
\bibitem [{\citenamefont {Richard}, \citenamefont {Wang},\ and\ \citenamefont
  {Zhao}(2015)}]{PhysRevC.91.014003}%
  \BibitemOpen
  \bibfield  {author} {\bibinfo {author} {\bibfnamefont {J.-M.}\ \bibnamefont
  {Richard}}, \bibinfo {author} {\bibfnamefont {Q.}~\bibnamefont {Wang}}, \
  and\ \bibinfo {author} {\bibfnamefont {Q.}~\bibnamefont {Zhao}},\ }\href
  {\doibase 10.1103/PhysRevC.91.014003} {\bibfield  {journal} {\bibinfo
  {journal} {Phys. Rev. C}\ }\textbf {\bibinfo {volume} {91}},\ p.\ \bibinfo
  {pages} {014003} (\bibinfo {year} {2015})}\BibitemShut {NoStop}%
\bibitem [{\citenamefont {Ahn}\ \emph {et~al.}(2001)\citenamefont {Ahn} \emph
  {et~al.}}]{PhysRevLett.87.132504}%
  \BibitemOpen
  \bibfield  {author} {\bibinfo {author} {\bibfnamefont {J.~K.}\ \bibnamefont
  {Ahn}} \emph {et~al.},\ }\href {\doibase 10.1103/PhysRevLett.87.132504}
  {\bibfield  {journal} {\bibinfo  {journal} {Phys. Rev. Lett.}\ }\textbf
  {\bibinfo {volume} {87}},\ p.\ \bibinfo {pages} {132504} (\bibinfo {year}
  {2001})}\BibitemShut {NoStop}%
\bibitem [{\citenamefont {Botvina}\ and\ \citenamefont
  {Pochodzalla}(2007)}]{PhysRevC.76.024909}%
  \BibitemOpen
  \bibfield  {author} {\bibinfo {author} {\bibfnamefont {A.}~\bibnamefont
  {Botvina}}\ and\ \bibinfo {author} {\bibfnamefont {J.}~\bibnamefont
  {Pochodzalla}},\ }\href {\doibase 10.1103/PhysRevC.76.024909} {\bibfield
  {journal} {\bibinfo  {journal} {Phys. Rev. C}\ }\textbf {\bibinfo {volume}
  {76}},\ p.\ \bibinfo {pages} {024909}Aug (\bibinfo {year}
  {2007})}\BibitemShut {NoStop}%
\bibitem [{\citenamefont {Lorente}, \citenamefont {Botvina},\ and\
  \citenamefont {Pochodzalla}(2011)}]{Lorente2011222}%
  \BibitemOpen
  \bibfield  {author} {\bibinfo {author} {\bibfnamefont {A.~S.}\ \bibnamefont
  {Lorente}}, \bibinfo {author} {\bibfnamefont {A.~S.}\ \bibnamefont
  {Botvina}}, \ and\ \bibinfo {author} {\bibfnamefont {J.}~\bibnamefont
  {Pochodzalla}},\ }\href {\doibase
  http://dx.doi.org/10.1016/j.physletb.2011.02.002} {\bibfield  {journal}
  {\bibinfo  {journal} {Physics Letters B}\ }\textbf {\bibinfo {volume}
  {697}},\ \unskip\ \bibinfo {pages} {222 -- 228} (\bibinfo {year}
  {2011})}\BibitemShut {NoStop}%
\bibitem [{\citenamefont {Batty}, \citenamefont {Friedman},\ and\ \citenamefont
  {Gal}(1999)}]{PhysRevC.59.295}%
  \BibitemOpen
  \bibfield  {author} {\bibinfo {author} {\bibfnamefont {C.~J.}\ \bibnamefont
  {Batty}}, \bibinfo {author} {\bibfnamefont {E.}~\bibnamefont {Friedman}}, \
  and\ \bibinfo {author} {\bibfnamefont {A.}~\bibnamefont {Gal}},\ }\href
  {\doibase 10.1103/PhysRevC.59.295} {\bibfield  {journal} {\bibinfo  {journal}
  {Phys. Rev. C}\ }\textbf {\bibinfo {volume} {59}},\ \unskip\ \bibinfo {pages}
  {295--304} (\bibinfo {year} {1999})}\BibitemShut {NoStop}%
\bibitem [{\citenamefont {Dover}\ and\ \citenamefont
  {Gal}(1983)}]{DOVER1983309}%
  \BibitemOpen
  \bibfield  {author} {\bibinfo {author} {\bibfnamefont {C.}~\bibnamefont
  {Dover}}\ and\ \bibinfo {author} {\bibfnamefont {A.}~\bibnamefont {Gal}},\
  }\href {\doibase http://dx.doi.org/10.1016/0003-4916(83)90036-2} {\bibfield
  {journal} {\bibinfo  {journal} {Annals of Physics}\ }\textbf {\bibinfo
  {volume} {146}},\ \unskip\ \bibinfo {pages} {309 -- 348} (\bibinfo {year}
  {1983})}\BibitemShut {NoStop}%
\bibitem [{\citenamefont {Friedman}\ and\ \citenamefont
  {Gal}(2007)}]{Friedman200789}%
  \BibitemOpen
  \bibfield  {author} {\bibinfo {author} {\bibfnamefont {E.}~\bibnamefont
  {Friedman}}\ and\ \bibinfo {author} {\bibfnamefont {A.}~\bibnamefont {Gal}},\
  }\href {\doibase http://dx.doi.org/10.1016/j.physrep.2007.08.002} {\bibfield
  {journal} {\bibinfo  {journal} {Physics Reports}\ }\textbf {\bibinfo {volume}
  {452}},\ \unskip\ \bibinfo {pages} {89 -- 153} (\bibinfo {year}
  {2007})}\BibitemShut {NoStop}%
\bibitem [{\citenamefont {Lalazissis}, \citenamefont {Grypeos},\ and\
  \citenamefont {Massen}(1989)}]{0954-3899-15-3-008}%
  \BibitemOpen
  \bibfield  {author} {\bibinfo {author} {\bibfnamefont {G.}~\bibnamefont
  {Lalazissis}}, \bibinfo {author} {\bibfnamefont {M.}~\bibnamefont {Grypeos}},
  \ and\ \bibinfo {author} {\bibfnamefont {S.}~\bibnamefont {Massen}},\ }\href
  {http://stacks.iop.org/0954-3899/15/i=3/a=008} {\bibfield  {journal}
  {\bibinfo  {journal} {Journal of Physics G: Nuclear and Particle Physics}\
  }\textbf {\bibinfo {volume} {15}},\ p.\ \bibinfo {pages} {303} (\bibinfo
  {year} {1989})}\BibitemShut {NoStop}%
\bibitem [{\citenamefont {Khaustov}\ \emph {et~al.}(2000)\citenamefont
  {Khaustov} \emph {et~al.}}]{PhysRevC.61.054603}%
  \BibitemOpen
  \bibfield  {author} {\bibinfo {author} {\bibfnamefont {P.}~\bibnamefont
  {Khaustov}} \emph {et~al.} (\bibinfo {collaboration} {The AGS E885
  Collaboration}),\ }\href {\doibase 10.1103/PhysRevC.61.054603} {\bibfield
  {journal} {\bibinfo  {journal} {Phys. Rev. C}\ }\textbf {\bibinfo {volume}
  {61}},\ p.\ \bibinfo {pages} {054603} (\bibinfo {year} {2000})}\BibitemShut
  {NoStop}%
\bibitem [{\citenamefont {Pochodzalla}, \citenamefont {Botvina},\ and\
  \citenamefont {Sanchez~Lorente}(2010)}]{Pochodzalla:2010}%
  \BibitemOpen
  \bibfield  {author} {\bibinfo {author} {\bibfnamefont {J.}~\bibnamefont
  {Pochodzalla}}, \bibinfo {author} {\bibfnamefont {A.}~\bibnamefont
  {Botvina}}, \ and\ \bibinfo {author} {\bibfnamefont {A.}~\bibnamefont
  {Sanchez~Lorente}},\ }\enquote {\bibinfo {title} {Studies of hyperons and
  antihyperons in nuclei},}\ in\ \href {\doibase DOI:
  https://doi.org/10.22323/1.103.0033} {\emph {\bibinfo {booktitle} {XLVIII
  International Winter Meeting on Nuclear Physics (Bormio2010)}}}\ (\bibinfo
  {year} {2010})\BibitemShut {NoStop}%
\bibitem [{\citenamefont {Kumagai-Fuse}\ and\ \citenamefont
  {Okabe}(2002)}]{PhysRevC.66.014003}%
  \BibitemOpen
  \bibfield  {author} {\bibinfo {author} {\bibfnamefont {I.}~\bibnamefont
  {Kumagai-Fuse}}\ and\ \bibinfo {author} {\bibfnamefont {S.}~\bibnamefont
  {Okabe}},\ }\href {\doibase 10.1103/PhysRevC.66.014003} {\bibfield  {journal}
  {\bibinfo  {journal} {Phys. Rev. C}\ }\textbf {\bibinfo {volume} {66}},\ p.\
  \bibinfo {pages} {014003} (\bibinfo {year} {2002})}\BibitemShut {NoStop}%
\bibitem [{\citenamefont {Randeniya}\ and\ \citenamefont
  {Hungerford}(2007)}]{PhysRevC.76.064308}%
  \BibitemOpen
  \bibfield  {author} {\bibinfo {author} {\bibfnamefont {S.~D.}\ \bibnamefont
  {Randeniya}}\ and\ \bibinfo {author} {\bibfnamefont {E.~V.}\ \bibnamefont
  {Hungerford}},\ }\href {\doibase 10.1103/PhysRevC.76.064308} {\bibfield
  {journal} {\bibinfo  {journal} {Phys. Rev. C}\ }\textbf {\bibinfo {volume}
  {76}},\ p.\ \bibinfo {pages} {064308} (\bibinfo {year} {2007})}\BibitemShut
  {NoStop}%
\bibitem [{\citenamefont {Nakano}(2000)}]{phdthesisNakano2000}%
  \BibitemOpen
  \bibfield  {author} {\bibinfo {author} {\bibfnamefont {J.~P.}\ \bibnamefont
  {Nakano}},\ }\enquote {\bibinfo {title} {Study of double~lambda hypernuclei
  by using cylindrical detector system},}\ \href@noop {} {Ph.D. thesis},\
  \bibinfo  {school} {Center for Nuclear Study, Graduate School of Science, the
  University of Tokyo}12 \bibinfo {year} {2000},\ \bibinfo {note} {cNS Report
  CNS-REP-28}\BibitemShut {NoStop}%
\bibitem [{\citenamefont {Mayeur}\ \emph {et~al.}(2016)\citenamefont {Mayeur},
  \citenamefont {Sacton}, \citenamefont {Vilain}, \citenamefont {Wilquet},
  \citenamefont {Stanley}, \citenamefont {Allen}, \citenamefont {Davis},
  \citenamefont {Fletcher}, \citenamefont {Garbutt}, \citenamefont {Shaukat},
  \citenamefont {Allen}, \citenamefont {Bull}, \citenamefont {Conway},\ and\
  \citenamefont {March}}]{Mayeur2016}%
  \BibitemOpen
  \bibfield  {author} {\bibinfo {author} {\bibfnamefont {C.}~\bibnamefont
  {Mayeur}}, \bibinfo {author} {\bibfnamefont {J.}~\bibnamefont {Sacton}},
  \bibinfo {author} {\bibfnamefont {P.}~\bibnamefont {Vilain}}, \bibinfo
  {author} {\bibfnamefont {G.}~\bibnamefont {Wilquet}}, \bibinfo {author}
  {\bibfnamefont {D.}~\bibnamefont {Stanley}}, \bibinfo {author} {\bibfnamefont
  {P.}~\bibnamefont {Allen}}, \bibinfo {author} {\bibfnamefont {D.~H.}\
  \bibnamefont {Davis}}, \bibinfo {author} {\bibfnamefont {E.~R.}\ \bibnamefont
  {Fletcher}}, \bibinfo {author} {\bibfnamefont {D.~A.}\ \bibnamefont
  {Garbutt}}, \bibinfo {author} {\bibfnamefont {M.~A.}\ \bibnamefont
  {Shaukat}}, \bibinfo {author} {\bibfnamefont {J.~E.}\ \bibnamefont {Allen}},
  \bibinfo {author} {\bibfnamefont {V.~A.}\ \bibnamefont {Bull}}, \bibinfo
  {author} {\bibfnamefont {A.~P.}\ \bibnamefont {Conway}}, \ and\ \bibinfo
  {author} {\bibfnamefont {P.~V.}\ \bibnamefont {March}},\ }\href {\doibase
  10.1007/BF02753195} {\bibfield  {journal} {\bibinfo  {journal} {Il Nuovo
  Cimento A (1971-1996)}\ }\textbf {\bibinfo {volume} {43}},\ p.\ \bibinfo
  {pages} {180} (\bibinfo {year} {2016})}\BibitemShut {NoStop}%
\bibitem [{\citenamefont {Gajewski}\ \emph {et~al.}(1967)\citenamefont
  {Gajewski}, \citenamefont {Mayeur}, \citenamefont {Sacton}, \citenamefont
  {Vilain}, \citenamefont {Wilquet}, \citenamefont {Harmsen}, \citenamefont
  {Setti}, \citenamefont {Raymund}, \citenamefont {Zakrzewski}, \citenamefont
  {Stanley}, \citenamefont {Davis}, \citenamefont {Fletcher}, \citenamefont
  {Allen}, \citenamefont {Bull}, \citenamefont {Conway},\ and\ \citenamefont
  {March}}]{Gajewski1967105}%
  \BibitemOpen
  \bibfield  {author} {\bibinfo {author} {\bibfnamefont {W.}~\bibnamefont
  {Gajewski}}, \bibinfo {author} {\bibfnamefont {C.}~\bibnamefont {Mayeur}},
  \bibinfo {author} {\bibfnamefont {J.}~\bibnamefont {Sacton}}, \bibinfo
  {author} {\bibfnamefont {P.}~\bibnamefont {Vilain}}, \bibinfo {author}
  {\bibfnamefont {G.}~\bibnamefont {Wilquet}}, \bibinfo {author} {\bibfnamefont
  {D.}~\bibnamefont {Harmsen}}, \bibinfo {author} {\bibfnamefont {R.~L.}\
  \bibnamefont {Setti}}, \bibinfo {author} {\bibfnamefont {M.}~\bibnamefont
  {Raymund}}, \bibinfo {author} {\bibfnamefont {J.}~\bibnamefont {Zakrzewski}},
  \bibinfo {author} {\bibfnamefont {D.}~\bibnamefont {Stanley}}, \bibinfo
  {author} {\bibfnamefont {D.~H.}\ \bibnamefont {Davis}}, \bibinfo {author}
  {\bibfnamefont {E.~R.}\ \bibnamefont {Fletcher}}, \bibinfo {author}
  {\bibfnamefont {J.~E.}\ \bibnamefont {Allen}}, \bibinfo {author}
  {\bibfnamefont {V.~A.}\ \bibnamefont {Bull}}, \bibinfo {author}
  {\bibfnamefont {A.~P.}\ \bibnamefont {Conway}}, \ and\ \bibinfo {author}
  {\bibfnamefont {P.~V.}\ \bibnamefont {March}},\ }\href {\doibase DOI:
  10.1016/0550-3213(67)90095-8} {\bibfield  {journal} {\bibinfo  {journal}
  {Nuclear Physics B}\ }\textbf {\bibinfo {volume} {1}},\ \unskip\ \bibinfo
  {pages} {105 -- 113} (\bibinfo {year} {1967})}\BibitemShut {NoStop}%
\bibitem [{\citenamefont {Juric}\ \emph {et~al.}(1973)\citenamefont {Juric},
  \citenamefont {Bohm}, \citenamefont {Klabuhn}, \citenamefont {Krecker},
  \citenamefont {Wysotzki}, \citenamefont {Coremans-Bertrand}, \citenamefont
  {Sacton}, \citenamefont {Wilquet}, \citenamefont {Cantwell}, \citenamefont
  {Esmael}, \citenamefont {Montwill}, \citenamefont {Davis}, \citenamefont
  {Kielczewska}, \citenamefont {Pniewski}, \citenamefont {Tymieniecka},\ and\
  \citenamefont {Zakrzewski}}]{JURIC19731}%
  \BibitemOpen
  \bibfield  {author} {\bibinfo {author} {\bibfnamefont {M.}~\bibnamefont
  {Juric}}, \bibinfo {author} {\bibfnamefont {G.}~\bibnamefont {Bohm}},
  \bibinfo {author} {\bibfnamefont {J.}~\bibnamefont {Klabuhn}}, \bibinfo
  {author} {\bibfnamefont {U.}~\bibnamefont {Krecker}}, \bibinfo {author}
  {\bibfnamefont {F.}~\bibnamefont {Wysotzki}}, \bibinfo {author}
  {\bibfnamefont {G.}~\bibnamefont {Coremans-Bertrand}}, \bibinfo {author}
  {\bibfnamefont {J.}~\bibnamefont {Sacton}}, \bibinfo {author} {\bibfnamefont
  {G.}~\bibnamefont {Wilquet}}, \bibinfo {author} {\bibfnamefont
  {T.}~\bibnamefont {Cantwell}}, \bibinfo {author} {\bibfnamefont
  {F.}~\bibnamefont {Esmael}}, \bibinfo {author} {\bibfnamefont
  {A.}~\bibnamefont {Montwill}}, \bibinfo {author} {\bibfnamefont
  {D.}~\bibnamefont {Davis}}, \bibinfo {author} {\bibfnamefont
  {D.}~\bibnamefont {Kielczewska}}, \bibinfo {author} {\bibfnamefont
  {T.}~\bibnamefont {Pniewski}}, \bibinfo {author} {\bibfnamefont
  {T.}~\bibnamefont {Tymieniecka}}, \ and\ \bibinfo {author} {\bibfnamefont
  {J.}~\bibnamefont {Zakrzewski}},\ }\href {\doibase
  http://dx.doi.org/10.1016/0550-3213(73)90084-9} {\bibfield  {journal}
  {\bibinfo  {journal} {Nuclear Physics B}\ }\textbf {\bibinfo {volume} {52}},\
  \unskip\ \bibinfo {pages} {1 -- 30} (\bibinfo {year} {1973})}\BibitemShut
  {NoStop}%
\bibitem [{\citenamefont {Breivik}\ \emph {et~al.}(1959)\citenamefont
  {Breivik}, \citenamefont {Skjeggestad}, \citenamefont {S{\"o}rensen},\ and\
  \citenamefont {Solheim}}]{Breivik1959}%
  \BibitemOpen
  \bibfield  {author} {\bibinfo {author} {\bibfnamefont {F.}~\bibnamefont
  {Breivik}}, \bibinfo {author} {\bibfnamefont {O.}~\bibnamefont
  {Skjeggestad}}, \bibinfo {author} {\bibfnamefont {S.~O.}\ \bibnamefont
  {S{\"o}rensen}}, \ and\ \bibinfo {author} {\bibfnamefont {A.}~\bibnamefont
  {Solheim}},\ }\href {\doibase 10.1007/BF02745755} {\bibfield  {journal}
  {\bibinfo  {journal} {Il Nuovo Cimento (1955-1965)}\ }\textbf {\bibinfo
  {volume} {12}},\ \unskip\ \bibinfo {pages} {531--540} (\bibinfo {year}
  {1959})}\BibitemShut {NoStop}%
\bibitem [{\citenamefont {Sacton}(1960)}]{Sacton1960}%
  \BibitemOpen
  \bibfield  {author} {\bibinfo {author} {\bibfnamefont {J.}~\bibnamefont
  {Sacton}},\ }\href {\doibase 10.1007/BF02860335} {\bibfield  {journal}
  {\bibinfo  {journal} {Il Nuovo Cimento (1955-1965)}\ }\textbf {\bibinfo
  {volume} {15}},\ \unskip\ \bibinfo {pages} {110--120} (\bibinfo {year}
  {1960})}\BibitemShut {NoStop}%
\bibitem [{\citenamefont {Kamada}\ \emph {et~al.}(1998)\citenamefont {Kamada},
  \citenamefont {Golak}, \citenamefont {Miyagawa}, \citenamefont {Wita\l{}a},\
  and\ \citenamefont {Gl\"ockle}}]{PhysRevC.57.1595}%
  \BibitemOpen
  \bibfield  {author} {\bibinfo {author} {\bibfnamefont {H.}~\bibnamefont
  {Kamada}}, \bibinfo {author} {\bibfnamefont {J.}~\bibnamefont {Golak}},
  \bibinfo {author} {\bibfnamefont {K.}~\bibnamefont {Miyagawa}}, \bibinfo
  {author} {\bibfnamefont {H.}~\bibnamefont {Wita\l{}a}}, \ and\ \bibinfo
  {author} {\bibfnamefont {W.}~\bibnamefont {Gl\"ockle}},\ }\href {\doibase
  10.1103/PhysRevC.57.1595} {\bibfield  {journal} {\bibinfo  {journal} {Phys.
  Rev. C}\ }\textbf {\bibinfo {volume} {57}},\ \unskip\ \bibinfo {pages}
  {1595--1603} (\bibinfo {year} {1998})}\BibitemShut {NoStop}%
\bibitem [{\citenamefont {Outa}\ \emph {et~al.}(1998)\citenamefont {Outa},
  \citenamefont {Aoki}, \citenamefont {Hayano}, \citenamefont {Ishikawa},
  \citenamefont {Iwasaki}, \citenamefont {Sakaguchi}, \citenamefont {Takada},
  \citenamefont {Tamura},\ and\ \citenamefont {Yamazaki}}]{OUTA1998251c}%
  \BibitemOpen
  \bibfield  {author} {\bibinfo {author} {\bibfnamefont {H.}~\bibnamefont
  {Outa}}, \bibinfo {author} {\bibfnamefont {M.}~\bibnamefont {Aoki}}, \bibinfo
  {author} {\bibfnamefont {R.}~\bibnamefont {Hayano}}, \bibinfo {author}
  {\bibfnamefont {T.}~\bibnamefont {Ishikawa}}, \bibinfo {author}
  {\bibfnamefont {M.}~\bibnamefont {Iwasaki}}, \bibinfo {author} {\bibfnamefont
  {A.}~\bibnamefont {Sakaguchi}}, \bibinfo {author} {\bibfnamefont
  {E.}~\bibnamefont {Takada}}, \bibinfo {author} {\bibfnamefont
  {H.}~\bibnamefont {Tamura}}, \ and\ \bibinfo {author} {\bibfnamefont
  {T.}~\bibnamefont {Yamazaki}},\ }\href {\doibase
  http://dx.doi.org/10.1016/S0375-9474(98)00281-4} {\bibfield  {journal}
  {\bibinfo  {journal} {Nuclear Physics A}\ }\textbf {\bibinfo {volume}
  {639}},\ \unskip\ \bibinfo {pages} {251c -- 260c} (\bibinfo {year}
  {1998})}\BibitemShut {NoStop}%
\bibitem [{\citenamefont {Bleser}(shed)}]{phdthesisBleser2018}%
  \BibitemOpen
  \bibfield  {author} {\bibinfo {author} {\bibfnamefont {S.}~\bibnamefont
  {Bleser}},\ }\enquote {\bibinfo {title} {{High resolution
  $\gamma$-spectroscopy of $\Lambda\Lambda$-hypernuclei at
  $\overline{\mbox{P}}$ANDA}},}\ \href@noop {} {Ph.D. thesis},\ \bibinfo
  {school} {Johannes Gutenberg-Universit{\"a}t Mainz} \bibinfo {year}
  {(unpublished)}\BibitemShut {NoStop}%
\bibitem [{\citenamefont {Gal}(2003)}]{Gal2003-private}%
  \BibitemOpen
  \bibfield  {author} {\bibinfo {author} {\bibfnamefont {A.}~\bibnamefont
  {Gal}},\ }\href@noop {} {}\ \bibinfo {howpublished} {Private communication} (
  \bibinfo {year} {2003} \unskip)\BibitemShut {NoStop}%
\bibitem [{\citenamefont {Pile}(2003)}]{HYP2003-Pile}%
  \BibitemOpen
  \bibfield  {author} {\bibinfo {author} {\bibfnamefont {P.}~\bibnamefont
  {Pile}},\ }\href@noop {} {\bibinfo {title} {{\textnormal{``Production of
  $\Lambda\Lambda$ Hypernuclei at the AGS''}}},\ }\ \bibinfo {howpublished}
  {{https://www.jlab.org/HYP2003/talks/Pile.pdf}} ( \bibinfo {year} {(2003)}
  \unskip),\ \bibinfo {note} {{\em{Talk given at VIII International Conference
  on Hypernuclear and Strange Particle Physics (HYP2003)}}}\BibitemShut
  {NoStop}%
\bibitem [{\citenamefont {Buss}\ \emph {et~al.}(2012)\citenamefont {Buss},
  \citenamefont {Gaitanos}, \citenamefont {Gallmeister}, \citenamefont {van
  Hees}, \citenamefont {Kaskulov}, \citenamefont {Lalakulich}, \citenamefont
  {Larionov}, \citenamefont {Leitner}, \citenamefont {Weil},\ and\
  \citenamefont {Mosel}}]{Buss20121}%
  \BibitemOpen
  \bibfield  {author} {\bibinfo {author} {\bibfnamefont {O.}~\bibnamefont
  {Buss}}, \bibinfo {author} {\bibfnamefont {T.}~\bibnamefont {Gaitanos}},
  \bibinfo {author} {\bibfnamefont {K.}~\bibnamefont {Gallmeister}}, \bibinfo
  {author} {\bibfnamefont {H.}~\bibnamefont {van Hees}}, \bibinfo {author}
  {\bibfnamefont {M.}~\bibnamefont {Kaskulov}}, \bibinfo {author}
  {\bibfnamefont {O.}~\bibnamefont {Lalakulich}}, \bibinfo {author}
  {\bibfnamefont {A.}~\bibnamefont {Larionov}}, \bibinfo {author}
  {\bibfnamefont {T.}~\bibnamefont {Leitner}}, \bibinfo {author} {\bibfnamefont
  {J.}~\bibnamefont {Weil}}, \ and\ \bibinfo {author} {\bibfnamefont
  {U.}~\bibnamefont {Mosel}},\ }\href {\doibase
  http://dx.doi.org/10.1016/j.physrep.2011.12.001} {\bibfield  {journal}
  {\bibinfo  {journal} {Physics Reports}\ }\textbf {\bibinfo {volume} {512}},\
  \unskip\ \bibinfo {pages} {1 -- 124} (\bibinfo {year} {2012})}\BibitemShut
  {NoStop}%
\bibitem [{\citenamefont {Gaitanos}\ and\ \citenamefont
  {Lenske}(2014)}]{GAITANOS2014256}%
  \BibitemOpen
  \bibfield  {author} {\bibinfo {author} {\bibfnamefont {T.}~\bibnamefont
  {Gaitanos}}\ and\ \bibinfo {author} {\bibfnamefont {H.}~\bibnamefont
  {Lenske}},\ }\href {\doibase https://doi.org/10.1016/j.physletb.2014.08.056}
  {\bibfield  {journal} {\bibinfo  {journal} {Physics Letters B}\ }\textbf
  {\bibinfo {volume} {737}},\ \unskip\ \bibinfo {pages} {256 -- 261} (\bibinfo
  {year} {2014})}\BibitemShut {NoStop}%
\bibitem [{\citenamefont {Rijken}\ and\ \citenamefont
  {Yamamoto}(2006)}]{Rijken:2006kg}%
  \BibitemOpen
  \bibfield  {author} {\bibinfo {author} {\bibfnamefont {T.~A.}\ \bibnamefont
  {Rijken}}\ and\ \bibinfo {author} {\bibfnamefont {Y.}~\bibnamefont
  {Yamamoto}},\ }\href@noop {} {\bibfield  {journal} {\bibinfo  {journal}
  {arXiv:nucl-th/0608074}\ } (\bibinfo {year} {2006})},\ \Eprint
  {http://arxiv.org/abs/nucl-th/0608074} {arXiv:nucl-th/0608074 [nucl-th]}
  \BibitemShut {NoStop}%
\bibitem [{\citenamefont {Rijken}, \citenamefont {Nagels},\ and\ \citenamefont
  {Yamamoto}(2010)}]{doi:10.1143/PTPS.185.14}%
  \BibitemOpen
  \bibfield  {author} {\bibinfo {author} {\bibfnamefont {T.~A.}\ \bibnamefont
  {Rijken}}, \bibinfo {author} {\bibfnamefont {M.~M.}\ \bibnamefont {Nagels}},
  \ and\ \bibinfo {author} {\bibfnamefont {Y.}~\bibnamefont {Yamamoto}},\
  }\href {\doibase 10.1143/PTPS.185.14} {\bibfield  {journal} {\bibinfo
  {journal} {Progress of Theoretical Physics Supplement}\ }\textbf {\bibinfo
  {volume} {185}},\ p.~\bibinfo {pages} {14} (\bibinfo {year}
  {2010})}\BibitemShut {NoStop}%
\bibitem [{\citenamefont {Fujiwara}\ \emph {et~al.}(2001)\citenamefont
  {Fujiwara}, \citenamefont {Kohno}, \citenamefont {Nakamoto},\ and\
  \citenamefont {Suzuki}}]{PhysRevC.64.054001}%
  \BibitemOpen
  \bibfield  {author} {\bibinfo {author} {\bibfnamefont {Y.}~\bibnamefont
  {Fujiwara}}, \bibinfo {author} {\bibfnamefont {M.}~\bibnamefont {Kohno}},
  \bibinfo {author} {\bibfnamefont {C.}~\bibnamefont {Nakamoto}}, \ and\
  \bibinfo {author} {\bibfnamefont {Y.}~\bibnamefont {Suzuki}},\ }\href
  {\doibase 10.1103/PhysRevC.64.054001} {\bibfield  {journal} {\bibinfo
  {journal} {Phys. Rev. C}\ }\textbf {\bibinfo {volume} {64}},\ p.\ \bibinfo
  {pages} {054001} (\bibinfo {year} {2001})}\BibitemShut {NoStop}%
\bibitem [{\citenamefont {Yamada}\ and\ \citenamefont
  {Ikeda}(1995)}]{YAMADA199579}%
  \BibitemOpen
  \bibfield  {author} {\bibinfo {author} {\bibfnamefont {T.}~\bibnamefont
  {Yamada}}\ and\ \bibinfo {author} {\bibfnamefont {K.}~\bibnamefont {Ikeda}},\
  }\href {\doibase https://doi.org/10.1016/0375-9474(94)00547-Z} {\bibfield
  {journal} {\bibinfo  {journal} {Nuclear Physics A}\ }\textbf {\bibinfo
  {volume} {585}},\ \unskip\ \bibinfo {pages} {79 -- 82} (\bibinfo {year}
  {1995})}\BibitemShut {NoStop}%
\bibitem [{\citenamefont {Yamada}\ and\ \citenamefont
  {Ikeda}(1997)}]{PhysRevC.56.3216}%
  \BibitemOpen
  \bibfield  {author} {\bibinfo {author} {\bibfnamefont {T.}~\bibnamefont
  {Yamada}}\ and\ \bibinfo {author} {\bibfnamefont {K.}~\bibnamefont {Ikeda}},\
  }\href {\doibase 10.1103/PhysRevC.56.3216} {\bibfield  {journal} {\bibinfo
  {journal} {Phys. Rev. C}\ }\textbf {\bibinfo {volume} {56}},\ \unskip\
  \bibinfo {pages} {3216--3230} (\bibinfo {year} {1997})}\BibitemShut {NoStop}%
\bibitem [{\citenamefont {Yamada}(1998)}]{YAMADA1998385c}%
  \BibitemOpen
  \bibfield  {author} {\bibinfo {author} {\bibfnamefont {T.}~\bibnamefont
  {Yamada}},\ }\href {\doibase https://doi.org/10.1016/S0375-9474(98)00301-7}
  {\bibfield  {journal} {\bibinfo  {journal} {Nuclear Physics A}\ }\textbf
  {\bibinfo {volume} {639}},\ \unskip\ \bibinfo {pages} {385c -- 388c}
  (\bibinfo {year} {1998})}\BibitemShut {NoStop}%
\bibitem [{\citenamefont {Hirataa}\ \emph {et~al.}(1998)\citenamefont
  {Hirataa}, \citenamefont {Nara}, \citenamefont {Ohnishi}, \citenamefont
  {Harada},\ and\ \citenamefont {Randrup}}]{HIRATAA1998389c}%
  \BibitemOpen
  \bibfield  {author} {\bibinfo {author} {\bibfnamefont {Y.}~\bibnamefont
  {Hirataa}}, \bibinfo {author} {\bibfnamefont {Y.}~\bibnamefont {Nara}},
  \bibinfo {author} {\bibfnamefont {A.}~\bibnamefont {Ohnishi}}, \bibinfo
  {author} {\bibfnamefont {T.}~\bibnamefont {Harada}}, \ and\ \bibinfo {author}
  {\bibfnamefont {J.}~\bibnamefont {Randrup}},\ }\href {\doibase
  http://dx.doi.org/10.1016/S0375-9474(98)00302-9} {\bibfield  {journal}
  {\bibinfo  {journal} {Nuclear Physics A}\ }\textbf {\bibinfo {volume}
  {639}},\ \unskip\ \bibinfo {pages} {389c -- 392c} (\bibinfo {year}
  {1998})}\BibitemShut {NoStop}%
\bibitem [{\citenamefont {Hirata}\ \emph {et~al.}(1999)\citenamefont {Hirata},
  \citenamefont {Nara}, \citenamefont {Ohnishi}, \citenamefont {Harada},\ and\
  \citenamefont {Randrup}}]{doi:10.1143/PTP.102.89}%
  \BibitemOpen
  \bibfield  {author} {\bibinfo {author} {\bibfnamefont {Y.}~\bibnamefont
  {Hirata}}, \bibinfo {author} {\bibfnamefont {Y.}~\bibnamefont {Nara}},
  \bibinfo {author} {\bibfnamefont {A.}~\bibnamefont {Ohnishi}}, \bibinfo
  {author} {\bibfnamefont {T.}~\bibnamefont {Harada}}, \ and\ \bibinfo {author}
  {\bibfnamefont {J.}~\bibnamefont {Randrup}},\ }\href {\doibase
  10.1143/PTP.102.89} {\bibfield  {journal} {\bibinfo  {journal} {Progress of
  Theoretical Physics}\ }\textbf {\bibinfo {volume} {102}},\ p.~\bibinfo
  {pages} {89} (\bibinfo {year} {1999})}\BibitemShut {NoStop}%
\bibitem [{\citenamefont {Ikeda}\ \emph {et~al.}(1994)\citenamefont {Ikeda},
  \citenamefont {Fukuda}, \citenamefont {Motoba}, \citenamefont {Takahashi},\
  and\ \citenamefont {Yamamoto}}]{doi:10.1143/ptp/91.4.747}%
  \BibitemOpen
  \bibfield  {author} {\bibinfo {author} {\bibfnamefont {K.}~\bibnamefont
  {Ikeda}}, \bibinfo {author} {\bibfnamefont {T.}~\bibnamefont {Fukuda}},
  \bibinfo {author} {\bibfnamefont {T.}~\bibnamefont {Motoba}}, \bibinfo
  {author} {\bibfnamefont {M.}~\bibnamefont {Takahashi}}, \ and\ \bibinfo
  {author} {\bibfnamefont {Y.}~\bibnamefont {Yamamoto}},\ }\href {\doibase
  10.1143/ptp/91.4.747} {\bibfield  {journal} {\bibinfo  {journal} {Progress of
  Theoretical Physics}\ }\textbf {\bibinfo {volume} {91}},\ p.\ \bibinfo
  {pages} {747} (\bibinfo {year} {1994})}\BibitemShut {NoStop}%
\bibitem [{\citenamefont {Yamamoto}\ \emph {et~al.}(1994)\citenamefont
  {Yamamoto}, \citenamefont {Motoba}, \citenamefont {Fukuda}, \citenamefont
  {Takahashi},\ and\ \citenamefont {Ikeda}}]{doi:10.1143/ptp.117.281}%
  \BibitemOpen
  \bibfield  {author} {\bibinfo {author} {\bibfnamefont {Y.}~\bibnamefont
  {Yamamoto}}, \bibinfo {author} {\bibfnamefont {T.}~\bibnamefont {Motoba}},
  \bibinfo {author} {\bibfnamefont {T.}~\bibnamefont {Fukuda}}, \bibinfo
  {author} {\bibfnamefont {M.}~\bibnamefont {Takahashi}}, \ and\ \bibinfo
  {author} {\bibfnamefont {K.}~\bibnamefont {Ikeda}},\ }\href {\doibase
  10.1143/ptp.117.281} {\bibfield  {journal} {\bibinfo  {journal} {Progress of
  Theoretical Physics Supplement}\ }\textbf {\bibinfo {volume} {117}},\
  \unskip\ \bibinfo {pages} {281--306} (\bibinfo {year} {1994})}\BibitemShut
  {NoStop}%
\bibitem [{\citenamefont {Aoki}\ \emph {et~al.}(2009)\citenamefont {Aoki} \emph
  {et~al.}}]{AOKI2009191}%
  \BibitemOpen
  \bibfield  {author} {\bibinfo {author} {\bibfnamefont {S.}~\bibnamefont
  {Aoki}} \emph {et~al.},\ }\href {\doibase
  http://dx.doi.org/10.1016/j.nuclphysa.2009.07.005} {\bibfield  {journal}
  {\bibinfo  {journal} {Nuclear Physics A}\ }\textbf {\bibinfo {volume}
  {828}},\ \unskip\ \bibinfo {pages} {191 -- 232} (\bibinfo {year}
  {2009})}\BibitemShut {NoStop}%
\bibitem [{\citenamefont {Yoshida}\ \emph {et~al.}(2017)\citenamefont
  {Yoshida}, \citenamefont {Kinbara}, \citenamefont {Mishina}, \citenamefont
  {Nakazawa}, \citenamefont {Soe}, \citenamefont {Theint},\ and\ \citenamefont
  {Tint}}]{YOSHIDA201786}%
  \BibitemOpen
  \bibfield  {author} {\bibinfo {author} {\bibfnamefont {J.}~\bibnamefont
  {Yoshida}}, \bibinfo {author} {\bibfnamefont {S.}~\bibnamefont {Kinbara}},
  \bibinfo {author} {\bibfnamefont {A.}~\bibnamefont {Mishina}}, \bibinfo
  {author} {\bibfnamefont {K.}~\bibnamefont {Nakazawa}}, \bibinfo {author}
  {\bibfnamefont {M.}~\bibnamefont {Soe}}, \bibinfo {author} {\bibfnamefont
  {A.}~\bibnamefont {Theint}}, \ and\ \bibinfo {author} {\bibfnamefont
  {K.}~\bibnamefont {Tint}},\ }\href {\doibase
  https://doi.org/10.1016/j.nima.2016.11.044} {\bibfield  {journal} {\bibinfo
  {journal} {Nuclear Instruments and Methods in Physics Research Section A:
  Accelerators, Spectrometers, Detectors and Associated Equipment}\ }\textbf
  {\bibinfo {volume} {847}},\ \unskip\ \bibinfo {pages} {86 -- 92} (\bibinfo
  {year} {2017})}\BibitemShut {NoStop}%
\end{thebibliography}
\end{document}